\documentclass[notitlepage,prd,showkeys,preprintnumbers,floatfix,superscriptaddress,nofootinbib]{revtex4-1}
\usepackage{float}
\usepackage{amsmath,amssymb}
\usepackage{graphicx}
\usepackage{mathtools,slashed}
\usepackage{bm}
\usepackage{comment}
\usepackage{xcolor}
\usepackage{subfigure}
\usepackage{array}
\usepackage{multirow}
\usepackage{url}
\usepackage{color}
\usepackage[T1]{fontenc}
\usepackage[latin9]{inputenc}
\usepackage{graphicx}
\usepackage{esint}
\usepackage{hyperref}
\usepackage{atbegshi}
\usepackage{lipsum}

\usepackage{glossaries-extra}
\setabbreviationstyle[acronym]{long-short}
\glssetcategoryattribute{acronym}{nohyperfirst}{true}

\newcommand{\nn}[0]{\nonumber}

\newcommand{\M}[0]{{\mathcal{M}}}
\newcommand{\T}[0]{{\mathcal{T}}}
\newcommand{\K}[0]{{\mathcal{K}}}
\newcommand{\cH}[0]{{\mathcal{H}}}
\newcommand{\bH}[0]{{\mathbf{H}}}

\newacronym{CM}{CM}{center-of-mass}
\newcommand{\CM}[0]{\gls{CM}}
\newacronym{BS}{BS}{Bethe-Salpeter}
\newcommand{\BS}[0]{\gls{BS}}

\newcommand{\CompRev}[0]{Radyushkin:1996nd,Diehl:1998kh,Belitsky:2001ns}
\newcommand{\OneDimQC}[0]{Sutherland:1971ic,Sutherland:1971ep,Sutherland:1971kq,Sutherland:1971ks}
\newcommand{\OpticalMai}[0]{Agadjanov:2016mao}
\newcommand{\MLlit}[0]{Hansen:2014eka, Hansen:2015zga,Briceno:2017tce,Briceno:2018mlh,Briceno:2018aml}

\begin{document}

\preprint{\vbox{\hbox{JLAB-THY-20-3210} }}
\preprint{\vbox{\hbox{CERN-TH-2020-112} }}

\title{The role of boundary conditions in quantum computations of scattering observables}

\author{Ra\'ul A.~Brice\~no}
\email[]{rbriceno@jlab.org}
\affiliation{Thomas Jefferson National Accelerator Facility, 12000 Jefferson Avenue, Newport News, Virginia 23606, USA}
\affiliation{ Department of Physics, Old Dominion University, Norfolk, Virginia 23529, USA}

\author{Juan V. Guerrero}
\email[]{juanvg@jlab.org}
\affiliation{Thomas Jefferson National Accelerator Facility, 12000 Jefferson Avenue, Newport News, Virginia 23606, USA}
\affiliation{ Department of Physics, Old Dominion University, Norfolk, Virginia 23529, USA}

\author{Maxwell T. Hansen}
\email[]{maxwell.hansen@cern.ch}
\affiliation{Theoretical Physics Department, CERN, 1211 Geneva 23, Switzerland}

\author{Alexandru Sturzu}
\email[]{alexandru.sturzu17@ncf.edu}
\affiliation{Department of Physics, New College of Florida, 5800 Bay Shore Rd, Sarasota, FL 34243, USA}

\date{\today}

\begin{abstract}
Quantum computing may offer the opportunity to simulate strongly-interacting field theories, such as quantum chromodynamics, with physical time evolution. This would give access to Minkowski-signature correlators, in contrast to the Euclidean calculations routinely performed at present. However, as with present-day calculations, quantum computation strategies still require the restriction to a finite system size, including a finite, usually periodic, spatial volume. In this work, we investigate the consequences of this in the extraction of hadronic and Compton-like scattering amplitudes. Using the framework presented in {\bf Phys. Rev. D101 014509 (2020)}, we quantify the volume effects for various $1+1$D Minkowski-signature quantities and show that these can be a significant source of systematic uncertainty, even for volumes that are very large by the standards of present-day Euclidean calculations. We then present an improvement strategy, based in the fact that the finite volume has a reduced symmetry. This implies that kinematic points, which yield the same Lorentz invariants, may still be physically distinct in the finite-volume system. As we demonstrate, both numerically and analytically, averaging over such sets can significantly suppress the unwanted volume distortions and improve the extraction of the physical scattering amplitudes. 
\end{abstract}


\nopagebreak
\maketitle

\section{Introduction\label{sec:intro}}

Non-perturbative descriptions lie at the core of a variety of interesting physical systems, ranging from strong electromagnetic fields, to topological effects in condensed matter systems, to the possible meta-stability and decay of the Standard Model vacuum. Another prominent example is quantum chromodynamics (QCD), the fundamental theory of the strong nuclear force, for which the properties of the low-energy degrees of freedom (the hadrons) cannot be analytically related to the underlying quantum fields (the quarks and gluons). To provide reliable predictions for such non-perturbative systems, one must often rely on numerical calculations. In the case of QCD, the most rigorous and well-established methodology is to numerically estimate a discretized version of the path integral, using Monte Carlo importance sampling.

This technique, known as lattice QCD, relies on the analytic continuation of the path-integral to Euclidean signature, such that the integrand becomes sharply peaked and can be reliably sampled in the high-dimensional space of field configurations. Consequently, the resulting correlation functions do not always have an obvious relation to physical observables, expressed in terms of Minkowski-signature correlators. The mismatch is especially relevant for dynamical observables that are intrinsically related to the time evolution of a system, including scattering and decay amplitudes as well as conductivities, viscosities, and other parameters describing the evolution of QCD under extreme conditions.

This has motivated a surge of activity to develop novel techniques that may provide direct access to Minkowski correlation functions. These efforts broadly fall under two camps: The first is to consider new Monte Carlo techniques that allow for a sampling of highly oscillatory integrands, for example by identifying field redefinitions that reduce the oscillatory behavior while leaving the resulting integral unchanged. Such techniques have already proven to be useful; see for example the Monte Carlo study of real time dynamics published in Ref.~\cite{Alexandru:2016gsd}. The second proposal is to reformulate the problem into one that is suitable to quantum computing techniques. We point the reader to Ref.~\cite{Georgescu:2013oza} for a review on these ideas and Refs.~\cite{Jordan:2014tma, Jordan:2011ci, Jordan:2011ne, Davoudi:2019bhy, Kuno:2014npa,Martinez:2016yna,Mueller:2019qqj, Lamm:2019uyc, Kaplan:2018vnj, Kaplan:2017ccd, Gustafson:2019mpk, Marshall:2015mna} for recent applications. Indeed, in a recent whitepaper~\cite{Joo:2019byq}, USQCD has encouraged the lattice QCD community \emph{to embark on an effort to understand the potential of quantum computing and quantum information science for QCD calculations important to high-energy and nuclear physics of the future.}

In this work, we discuss prospects for extracting scattering amplitudes from Minkowski-signature, finite-volume correlation functions. We consider both $2 \to 2$ hadronic amplitudes and $1+\mathcal{J}\to 1+\mathcal{J}$ Compton-like amplitudes, in which a hadron scatters off an external current. As discussed in further detail in the following sections, the Compton-like amplitudes turn out to be conceptually simpler for the analysis that we consider here.

Both classes of amplitudes are well motivated by a broad range of phenomenology. For example, the high-energy and high-virtuality limits of Compton scattering can be used determine Parton Distribution Functions (PDFs) and Generalized Parton Distributions (GPDs). Such distributions are at the core of the present-day Jefferson Lab 12 GeV program as well as the Department of Energy's future \href{https://www.bnl.gov/eic/}{Electron Ion Collider (EIC)}. In addition, this same class of amplitudes may give access to inclusive neutrino-nucleon scattering, a pressing need for future analysis of Fermilab's \href{https://www.fnal.gov/pub/science/lbnf-dune/index.html}{Deep Underground Neutrino Experiment (DUNE)}.

\bigskip

Although many aspects of the detailed set-up differ between Euclidean-signature lattice calculations and Minkowski-signature quantum computations, both require a truncation of the single-particle Hilbert space, i.e.~a finite set of discrete momenta, in order to define a system that fits on a finite-sized computer. As a result, both approaches will give access to finite-volume correlation functions, which can be written as discrete sums involving finite-volume matrix elements and energies. As a specific example, we consider the Minkowski-signature finite-volume correlator most closely related to Compton scattering
\begin{equation}
\begin{split}
\label{eq:CLdef}
C^{\text{M}}_{\boldsymbol p, L}(t) & \equiv \langle \boldsymbol p, L \vert \, \text{T} \{ J^\dagger_{-\boldsymbol p}(t) \, J_{\boldsymbol p}(0) \} \, \vert \boldsymbol p, L \rangle \,, \\[5pt]
& = \sum_{n} c_{n}(\boldsymbol p, L) \, e^{- i [E_{n}(L) - E_{\boldsymbol p}(L)] t} \,, \qquad \qquad (t>0) \,,
\end{split} 
\end{equation}
where in the second step we have inserted a complete set of finite-volume states, assuming $t>0$. Here $c_{n}(\boldsymbol p, L)$ is defined as a product of matrix elements of the current $J_{\boldsymbol p}(0)$, the details of which are irrelevant. The time dependence is governed by the difference of $E_{\boldsymbol p}(L)$, the finite-volume energy of a single-particle state $\vert \boldsymbol p, L \rangle$ with momentum $\boldsymbol p$, and $E_{n}(L)$, the $n$th finite-volume excited-state energy with zero momentum. The analogous Euclidean correlation function is given by
\begin{equation}
\begin{split}
\label{eq:CLdef}
C^{\text{E}}_{\boldsymbol p, L}(\tau) = \sum_{n} c_{n}(\boldsymbol p, L) \, e^{- [E_{n}(L) - E_{\boldsymbol p}(L)] \tau} \,, \qquad \qquad (\tau>0) \,.
\end{split} 
\end{equation}

The key observation here is that exactly the same overlap factors $c_{n}(\boldsymbol p, L) $ and energy differences $E_{n}(L) - E_{\boldsymbol p}(L)$ enter the two correlation functions, with only the functional distinction of oscillating vs.~decay exponentials giving the difference. With this in mind, especially considering the significant investment in developing viable real-time computations, it is important to understand to what extent $C^{\text{M}}_{\boldsymbol p, L}(t)$ gives a more useful prediction as compared to $C^{\text{E}}_{\boldsymbol p, L}(t) $. In this work we argue that Euclidean-signature correlators may, in fact, be more useful in a certain range of kinematics. In addition we will show that $C^{\text{M}}_{\boldsymbol p, L}(t)$ can easily be dominated by finite-volume effects unless specific strategies are employed to estimate and remove these.

%
To give a meaningful comparison, it is important to benchmark future proposals against the current state of the art.
In this vein we note that the best-established approach, at present, is to numerically determine finite-volume energies $E_n(L)$ and matrix elements $c_n(\boldsymbol p, L)$ from Euclidean correlators and, by making use of model-independent field-theoretic relations, to map these into physically observable scattering and decay amplitudes. This strategy was pioneered by L\"uscher \cite{Luscher:1986pf,Luscher:1991n1} who derived a relation between finite-volume energies and the two-to-two scattering amplitude of identical scalar (or pseudo-scalar) particles. 

These techniques have reached a high level of maturity. On the formal side, L\"uscher's original work has since been generalized for any number of coupled two-particle channels, including non-identical and non-degenerate particles with any intrinsic spin~\cite{Rummukainen:1995vs, Kim:2005gf, He:2005ey, Davoudi:2011md, Hansen:2012tf, Briceno:2012yi, Briceno:2013lba, Briceno:2014oea}. This, in turn, has led to a wide class of phenomenologically interesting lattice QCD studies; see for example Refs.~\cite{Wilson:2015dqa, Briceno:2016mjc, Brett:2018jqw, Guo:2018zss, Andersen:2017una, Andersen:2018mau, Dudek:2014qha, Dudek:2016cru, Woss:2018irj, Woss:2019hse, Orginos:2015aya, Berkowitz:2015eaa}. The ideas have further been generalized for the study of electroweak process involving matrix elements of local currents with two-hadron asymptotic states~\cite{Lellouch:2000, Meyer:2011um, Briceno:2012yi, Briceno:2014uqa, Feng:2014gba, Briceno:2015csa} including two-to-two transitions mediated by an external current \cite{Briceno:2015tza, Baroni:2018iau} as well long-range matrix elements involving currents displaced in time \cite{Christ:2015pwa, Briceno:2019opb}.

All such results are exact up to volume corrections that scale as $e^{- m L}$, with $m$ the mass of the lightest degree of freedom, but are limited to energies lying below the lowest-lying threshold with three or more particles. Given the importance of this restriction, considerable effort has been invested recently in extending the framework to kinematics for which three-particle states can go on-shell~\cite{Polejaeva:2012ut, Briceno:2012rv, Hansen:2014eka, Hansen:2015zga, Hammer:2017uqm,Hammer:2017kms, Guo:2017ism,Mai:2017bge, Briceno:2017tce, Doring:2018xxx, Briceno:2018mlh, Mai:2018djl, Briceno:2018aml, Blanton:2019igq}. {See Refs.~\cite{Briceno:2017max} and \cite{Hansen:2019nir} for recent reviews on the status and implementation of the two- and three-particle finite-volume formalisms, respectively.} 

\bigskip

Despite the overwhelming success of these techniques for low to moderate energies, the development and implementation of such approaches becomes increasingly challenging as the energy is increased into the regime where multiple channels, especially those containing three or more particles, are open. As a result, it is useful to explore the possibility of extracting scattering amplitudes without using the finite-volume as a tool. One set of proposals in this direction involves estimating the inverse Laplace transform to recover a smeared version of either an inclusive total rate \cite{Hansen:2017mnd} or else a decay or scattering amplitude~\cite{Bulava:2019kbi}. In these methods the extracted quantity is distorted by finite-volume effects as well as the smearing width, which, in the case of the scattering amplitude, can be understood as a non-infinitesimal $i \epsilon$ prescription.

The distinction in a Minkowski calculation is that $C^{\text{M}}_{\boldsymbol p, L}(t)$ gives direct access to a finite-volume version of the rate or amplitude without the challenging inverse problem. Instead, one Fourier transforms the time coordinate to a corresponding energy, using an $i \epsilon$ prescription to regulate the large time behavior of the integral.  Denoting the result of this transformation by $\T_{L,\epsilon}(\omega, \boldsymbol p)$, where $\omega$ is the energy carried by the outgoing current, one must next estimate the ordered double limit\footnote{See also Ref.~\cite{Agadjanov:2016mao}, in which twisted boundary conditions are used to extract the finite-volume optical potential. Estimating the same ordered double limit on this object then allows one to extract the scattering amplitude of a single specified channel in the inelastic regime.}
\begin{equation}
\label{eq:orddoublim}
\T(P^2,  Q^2) = \lim_{\epsilon \to 0} \lim_{L \to \infty} \T_{L,\epsilon}(\omega, \boldsymbol p)  \,,
\end{equation}
where $P^2 = (\omega_{\boldsymbol p} + \omega)^2$ and $Q^2 = - \omega^2 + \boldsymbol p^2$, with $\omega_{\boldsymbol p} = \sqrt{m^2 + \boldsymbol p^2}$. The notation is meant to emphasize that the physical Compton amplitude, $\T$, only depends on two Lorentz invariants in the forward limit.

In this work, we explore the practical challenges of relating $C^{\text{M}}_{\boldsymbol p, L}(t)$ to $\T(P^2,  Q^2)$, as well as the analogous problem for hadronic amplitudes. Though the motivation differs, the underlying problem has overlap with Refs.~\cite{Hansen:2017mnd,Bulava:2019kbi} in which the same ordered double limit, Eq.~\eqref{eq:orddoublim}, was discussed in the context of the inverse Laplace transform. 
In this work we restrict attention to a 1+1D set-up as this is likely the first case to be studied by a quantum computer. In contrast to Refs.~\cite{Hansen:2017mnd,Bulava:2019kbi}, we numerically explore the Fourier transform of $C^{\text{M}}_{\boldsymbol p, L}(t)$ using the general finite-volume formalism presented in Ref.~\cite{Briceno:2019opb}. This allows one to predict the functional form of the finite-volume correlator, for a given set of infinite-volume amplitudes and matrix elements.

 The role of finite-$L$ in quantum computations of scattering was already discussed by Jordan, Lee and Preskill in Ref.~\cite{Jordan:2011ci}. In particular, the authors consider volume effects on the two-particle scattering amplitude, calculated in $\lambda \phi^4$ theory through $\mathcal O(\lambda^2)$, and argue that these are exponentially suppressed. However, this only holds in the Born approximation, which, as explained in Ref.~\cite{Jordan:2011ci}, amounts to neglecting the $s$-channel two particle loop [see Fig.~\ref{fig:iTL}(a)]. By contrast, the finite-volume formalism of L{\"u}scher and its various extensions are based on the observation that the $s$-channel loops generate the dominate finite-$L$ effects, and that these can be summed to all orders in perturbation theory, without making use of Born or non-relativistic approximations. The distinction explains why we reach qualitatively different conclusions about the importance of finite-volume effects in future quantum computations.

The remainder of this article is organized as follows: In the next section we review the relevant properties of infinite-volume hadronic and Compton-like amplitudes in 1+1D. Then, in Sec.~\ref{sec:FV}, we summarize the formalism of Ref.~\cite{Briceno:2019opb} for predicting the finite-volume versions of these quantities. Section~\ref{sec:numerical} contains the central new results of this work, which can be summarized as follows:
\begin{enumerate}
\item In Sec.~\ref{sec:fixedvolHad}, focusing on the hadronic amplitude, we demonstrate that, even for very large volumes ($mL=30$) and plausible choices of the infinite-volume inputs, cases arise in which it is impossible to identify a smearing width $\epsilon$ that gives a suitable estimate of the ordered double limit. 
\item In Sec.~\ref{sec:bin}, we turn to the Compton amplitude and describe strategies for estimating the infinite-volume limit with external kinematics held fixed. 
Here we show that averaging over distinct finite-volume kinematics, chosen to yield same Lorentz invariants and thus the same infinite-volume amplitudes, can dramatically reduce the finite-volume distortions.
\end{enumerate}
We also include two appendices. In Appendix \ref{app:ps} we give details on the relevant finite-volume function in 1+1D and and in Appendix \ref{app:averaging} we given an analytic explanation as to why momentum averaging suppresses finite-volume effects.

\section{Infinite-volume amplitudes in 1+1D\label{sec:IV_amps}}

In this section we review known properties of infinite-volume two-particle scattering amplitudes in 1+1D.

\bigskip

We begin with the properties of a two-body amplitude in the absence of external currents, in the energy regime $2 m < E^\star < 3m$, where $E^\star$ denotes the center of mass energy for the two-particle state. We denote the total energy and momentum in a general frame using the two-vector $P^\mu = (E, \boldsymbol P)$, where the bold symbol is used for the spatial part, even though this is a single-component in our 1+1D set-up. The usual kinematic relations hold
\begin{equation}
E^{\star2} = P^\mu P_\mu = E^2 - \boldsymbol P^2 = s \,,
\end{equation}
where we have introduced the Mandelstam variable, $s$, in the final equality. The restriction to $2 m < E^\star < 3m$ implies an energy regime where only elastic scattering occurs.

\begin{figure}[t!]
\begin{center}
\includegraphics[width=1\textwidth]{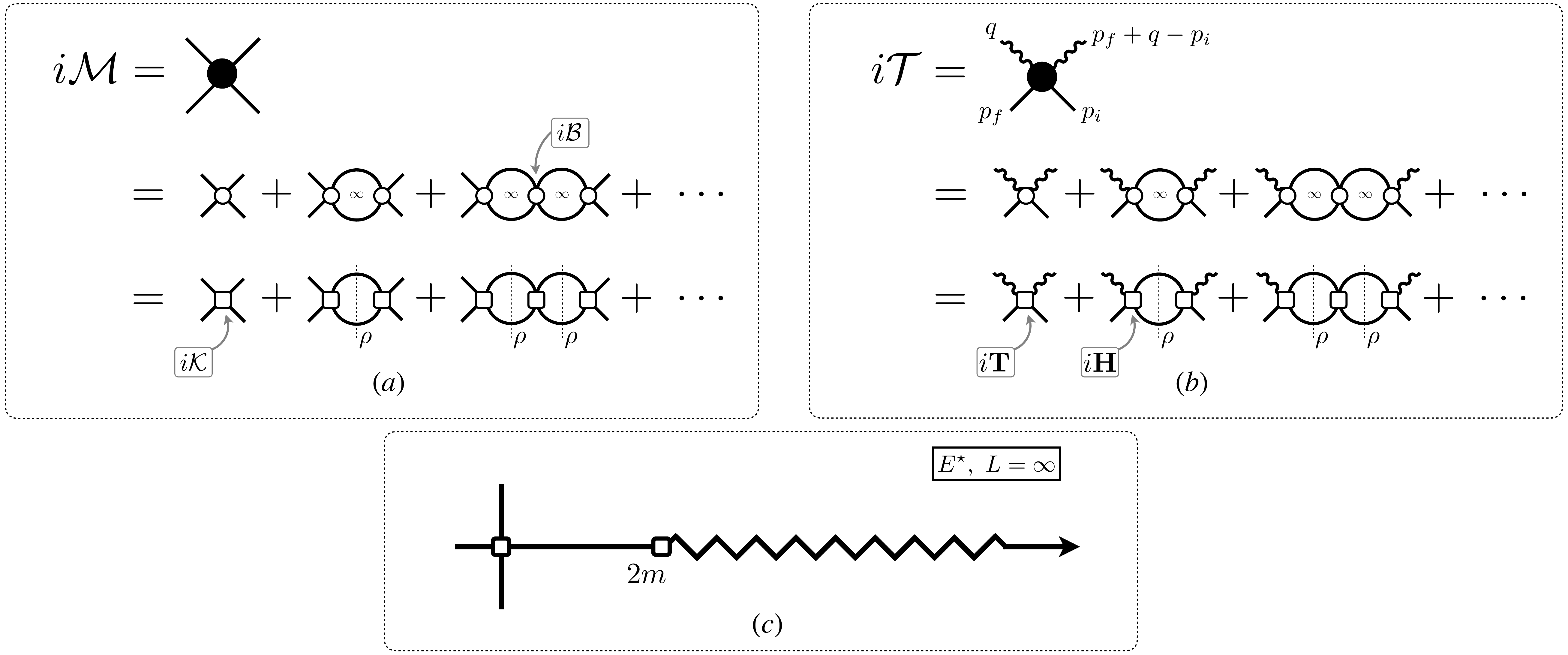}
\caption{Diagrammatic representations of $(a)$ the hadronic scattering amplitude, denoted by $\M$, and $(b)$ the Compton-like amplitude, denoted by $\T$. Panel $(c)$ shows the analytic structure of both $\M$ and $\T$ on the first Riemann Sheet.\label{fig:iM_iT}}
\end{center}
\end{figure}

\begin{figure}[t!]
\begin{center}
\includegraphics[width=.6\textwidth]{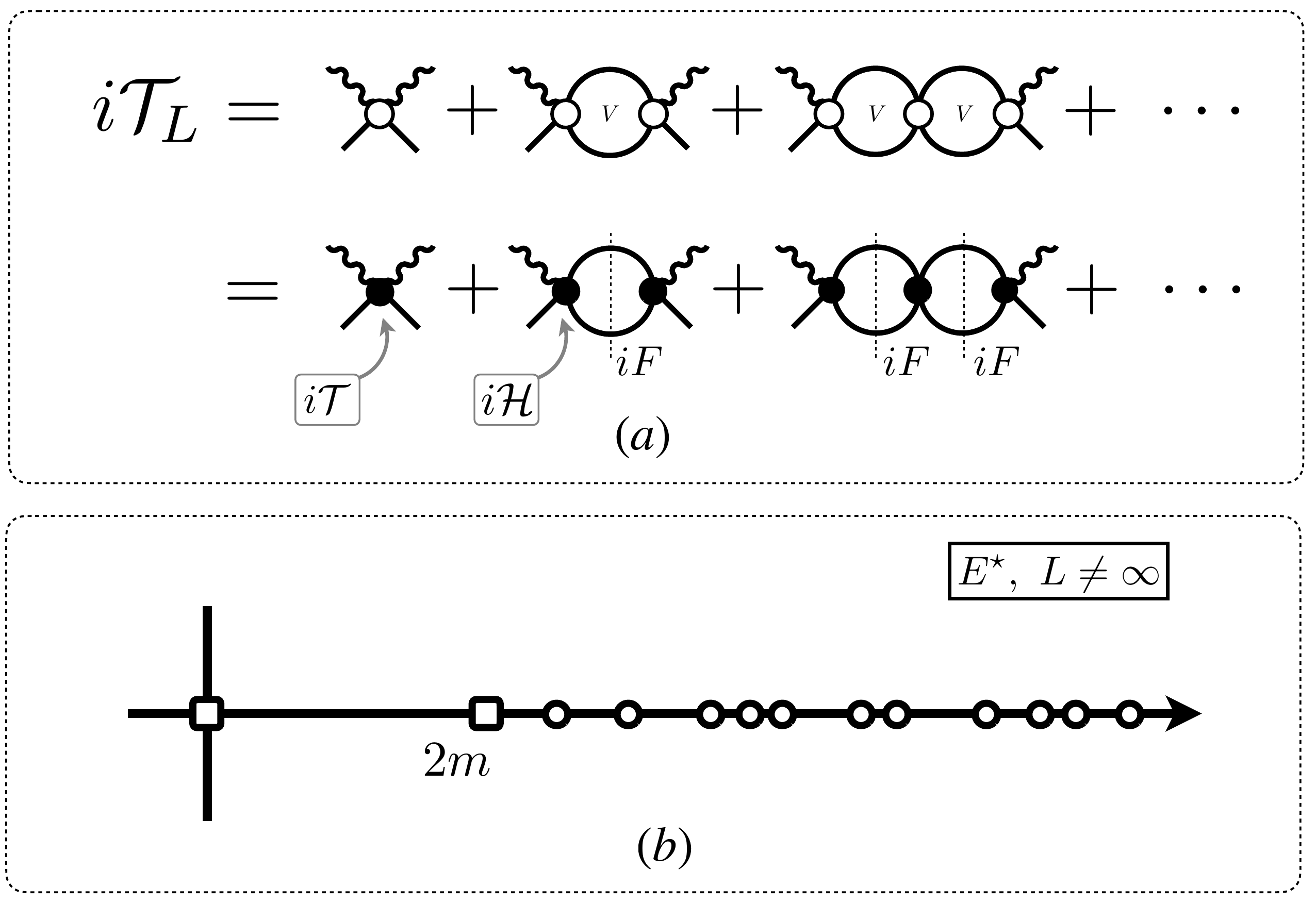}
\caption{$(a)$ Diagrammatic representation of the finite-volume Compton amplitude, denoted by $\T_L$. Panel $(b)$ shows the corresponding analytic structure, to be compared with Fig.~\ref{fig:iM_iT}$(c)$.}
\label{fig:iTL}
\end{center}
\end{figure}

In the second line of Fig.~\ref{fig:iM_iT}(a), we diagrammatically define the scattering amplitude, denoted by $\M(E^\star)$, as a sum of ladder diagrams, built from fully-dressed propagators and \BS~kernels. The \BS~kernels are defined as the sum of all Feynman diagrams with four external legs that are two-particle irreducible with respect to internal propagator sets carrying the total energy, $E^\star$. In other words, the \BS~kernels contain all diagrams that remain connected after any two lines, carrying the total energy, are cut. Combining this with all two-particle loops, as shown in Fig.~\ref{fig:iM_iT}(a), then leads to the proper inclusion of all diagrams.\footnote{The Born approximation, applied in this context in Ref.~\cite{Jordan:2011ci}, can be understood as approximating the scattering amplitude $\mathcal M$ by the \BS~kernel. In the language of non-relativistic quantum mechanics this corresponds to the Fourier transform of the potential.}

The utility of this expansion in the infinite-volume theory is that the \BS~kernels are real, meromorphic functions (analytic up to isolated poles) in a strip of the complex $s$ plane defined by $(2 m)^2 <\text{Re}[s] < (3 m)^2$. Thus, in the elastic regime, the complex-valued form of $\M(E^\star)$, as well as its non-analytic structure, arise only due to the two-particle loops shown explicitly. The diagrammatic expansion can be reduced by breaking each two-particle loop into a real-valued piece, defined via a principal-value pole prescription, together with the imaginary part. The latter leads to a phase-space factor, $\rho(E^\star)$, such that the series can be re-organized to give [see the third line of Fig.~\ref{fig:iM_iT}(a)]
\begin{equation}
\M(E^\star) = \sum_{n=0}^\infty \mathcal K(E^\star) \big [i \rho(E^\star) \mathcal K(E^\star) \big ]^n \,,
\end{equation}
where we have introduced the K matrix, $\mathcal K(E^\star)$, defined diagrammatically as the ladder-diagram series with principal-values in all two-particle loops. 
Here we have only displayed the dependence on $E^\star$, equivalently on the Mandelstam variable $s$. In the 1+1D theory the Mandelstam variables $t$ and $u$ can only taken on two discrete values $\{t, u\} = \{0, 4m^2 - s \}$ or else $\{t, u\} = \{4m^2 - s ,0\}$ corresponding to the two possible choices for the\!\!\! \CM~frame angle $\cos \theta = \pm 1$. The symmetric combinations of these two values is the 1+1D analog of the $S$-wave projection and the anti-symmetric the $P$-wave. The latter vanishes for identical particles, due to the $t \leftrightarrow u$ crossing symmetry, so we are left with a single scalar function of $\mathcal M(E^\star)$.
Summing the series leads to the compact result
\begin{align}
\M(E^\star)
=\frac{1}{\K(E^\star)^{-1}-i\rho(E^\star)} \,.
\label{eq:Mdef}
\end{align}

This expression, together with the fact that $\K(E^\star)$ is real for $2m < E^\star < 3m$, is equivalent to the constraint imposed by the unitarity of the S matrix.
To see this one combines the relation $S(E^\star) = 1+2i \rho(E^\star) \M(E^\star)$ with $S(E^\star)S(E^\star)^\dagger = \mathbb I$ to deduce
\begin{align}
\label{eq:unitaryM}
{\rm Im }\,\M(E^\star)
=\M^*(E^\star) \,\rho(E^\star)\,\M(E^\star) \,,
\end{align}
a constraining equation that is uniquely and generally solved by Eq.~\eqref{eq:Mdef}. In addition, unitarity in the single-channel sector requires $S(E^\star)=e^{i2\delta(E^\star)}$, for a real-valued scattering phase shift, $\delta(E^\star)$. This, in turn, implies the standard relation between scattering phase and K matrix:
\begin{align}
\K(E^\star)^{-1}= \rho(E^\star) \, \cot\delta(E^\star) \,.
\label{eq:cotd}
\end{align}

We stress that, while unitarity provides a more general, non-perturbative (indeed field-theory independent) basis for Eq.~\eqref{eq:Mdef}~\footnote{See Ref.~\cite{Jackura:2018xnx} for a recent example for constraints that unitarity places on three-body systems in 3+1D.}, the diagrammatic perspective is useful for two reasons. First, it leads to a simple expression for $\rho(E^\star)$ as the imaginary part of the two-particle loop. As we show in Appendix \ref{app:ps}, for a 1+1D relativistic scalar theory the result is
\begin{equation}
\rho(E^\star) = \frac{1 }{8 E^\star q^\star} \,,
\end{equation}
where $q^\star \equiv \sqrt{E^{\star 2}/4 - m^2}$, is the magnitude of a single particle's momentum in the \CM~frame. We comment that this is the source of the branch cut singularity depicted in Fig.~\ref{fig:iM_iT}($c$). Second, the diagrammatic perspective gives the extra constraint that $\mathcal K(E^\star)$ is not only real but in fact meromorphic in a strip about the real axis, in the regime of elastic scattering. This provides an important guide in parameterizing the scattering amplitude, for example by using the effective-range expansion.

\bigskip

We now turn our attention to Compton-like amplitudes, restricting our attention here to matrix elements of two scalar currents, denoted by $\mathcal J(x)$ and $\mathcal J'(x)$, between two single-particle external states 
\begin{equation}
\label{eq:TAdef}
\T(s,Q^2 ,Q^2_{if}) \equiv i \int d^2 x \, e^{i\omega t - i\bm{q} \cdot \bm{x}} \, \langle \boldsymbol p_f \vert \, \text{T} \{ \mathcal J (x) \mathcal J'(0) \} \, \vert \boldsymbol p_i \rangle_{\text{c}} \,,
\end{equation}
where T indicates time ordering, $q=(\omega, \bm{q})$, $ s = (p_f+q)^2$, $Q^2=-q^2$, $Q^2_{if}=-(p_f+q-p_i)^2$ and the subscript ``$\text{c}$'' means that only connected contributions are included in the definition of $\T$. 
As with the hadronic amplitude, here $t$ can take on two discrete values. But in contrast to $\mathcal M$ the $t \leftrightarrow u$ symmetry is broken and the anti-symmetric, $P$-wave-like projection is non-vanishing. For this work we simply restrict attention to $t=0$. For the other choice of $t$, the unitarity expressions given below are slightly complicated [see, again, Ref.~\cite{Briceno:2019opb}]. We do this in order to simplify the expressions below, but in practice the qualitative conclusions are expected to hold for arbitrary kinematics.

One can readily generalize these amplitudes by invoking any type of external states and local currents, including a generic Lorentz structure for the latter. For fully general expressions that follow the notation used here we point the reader to Ref.~\cite{Briceno:2019opb}, from which we take many key ideas. The importance of Compton-like amplitudes are discussed, for example, in Refs.~\cite{\CompRev}.

As discussed in Ref.~\cite{Briceno:2019opb}, $\T(s,Q^2 ,Q^2_{if})$ admits unitarity constraints that are closely related to those for $\mathcal M(s)$, summarized in Eqs.~\eqref{eq:Mdef} and \eqref{eq:unitaryM}. To derive these it is again useful to introduce a diagrammatic representation [see Fig.~\ref{fig:iM_iT}($b$)], built from fully-dressed propagators and \BS~kernels as well as new objects that do not arise in the decomposition of $\M(s)$. Specifically, three analogs of the \BS~kernel arise in which either \emph{(i)} one incoming particle, or \emph{(ii)} one outgoing particle, or else \emph{(iii)} one of each, is replaced by one of the two external currents. Following the same steps as with $\mathcal M(s)$ then leads to the result
\begin{align}
\label{eq:compton}
\T (E^\star,Q^2 ,Q^2_{if})&= \mathbf{T}(E^\star,Q^2 ,Q^2_{if}) + \bH(E^\star ,Q^2)
\frac{i}{1-i\rho(E^\star) \K(E^\star) }\rho(E^\star) \,
\bH' (E^\star ,Q^2_{if}) \,,
\end{align}
which is the Compton-amplitude generalization of Eq.~\eqref{eq:Mdef}. Here $ \mathbf{T}$, $\bH$ and $\bH' $ are modifications of the K matrix that appear due to the three new kernels [see Fig.~\ref{fig:iM_iT}($b$)]. They match $\K(E^\star)$ in their diagrammatic definitions, with the difference of external legs as shown in the last line of Fig.~\ref{fig:iM_iT}($b$). We point the reader to Ref.~\cite{Briceno:2019opb} for the detailed derivation of this result. To give a precise definition of $\bH$ and $\bH'$ it is useful to introduce another type of physical amplitude, the $1 + \mathcal J \to 2$ transition amplitude, denoted by $\mathcal H$. Then the following relations serve to define the bold quantities:
\begin{align}
\cH(E^\star,Q^2) &= \bH(E^\star,Q^2) \frac{1}{1-i \rho(E^\star) \,\K(E^\star) } \,, &
\label{eq:Hmatv2}
\cH'(E^\star,Q^2) &= \frac{1}{1-i \K(E^\star) \, \rho(E^\star) } \bH'(E^\star,Q^2) \,.
\end{align}

Equation~\eqref{eq:compton} is better understood by observing that, if the currents $\mathcal J(x)$ and $\mathcal J'(x)$ and the virtualities $Q^2$ and $Q^2_{if}$ can be chosen such that $ \mathbf{T}, \bH, \bH' \to \K(E^\star)$, then we recover
\begin{align}
\T (E^\star,Q^2 ,Q^2_{if}) \ \ \ \ \Longrightarrow \ \ \ \ \K(E^\star) + \K(E^\star) 
\frac{i}{1-i\rho(E^\star) \K(E^\star) }\rho(E^\star) \,
\K(E^\star) = \mathcal M(E^\star) \,.
\end{align}

In fact, as we now show, this operator choice can be realized in practice, such that the hadronic amplitude can be recovered from $\T (E^\star,Q^2 ,Q^2_{if})$. The key relation, which follows directly from the LSZ reduction formula, reads
\begin{align}
\label{eq:M_from_T}
\M (E^\star)&=
\lim_{Q^2 ,Q^2_{if}\to -m^2}
\frac{
(Q^2 + m^2)
(Q^2_{if} + m^2)}
{\langle 0 \vert \mathcal J(0) \vert q \rangle
\,
\langle \boldsymbol p_f+q-\boldsymbol p_i \vert \mathcal J'(0) \vert 0 \rangle}
\T (E^\star,Q^2 ,Q^2_{if}) \,,
\end{align}
where we have assumed that the currents $\mathcal J(0)$ and $\mathcal J'(0)$ overlap the single-particle states

We close this section with a final technical detail, discussed more thoroughly in Ref.~\cite{Briceno:2019opb}, that will be particularly important for our numerical studies in Sec.~\ref{sec:numerical}. The issue is associated with poles that can arise in the K matrix, $\mathcal K(E^\star)$. These occur in many physically realized systems, for example those with a narrow or Breit-Wigner-like resonance. In such cases, one can show from Eq.~\eqref{eq:Hmatv2}, and also from a diagrammatic analysis, that $\textbf H$ and $\textbf H'$ develop poles as well. Thus, in order to prevent unphysical poles from arising in the amplitude $\mathcal{T}(E^\star ,Q^2 ,Q^2_{if})$ one must require $\textbf T$ to take on the form
\begin{align}
\mathbf{T}(E^\star,Q^2 ,Q^2_{if}) = 
\bH(E^\star,Q^2) \, \frac{1}{\K(E^\star)}\,
\bH'(E^\star,Q^2_{if}) +\mathbf{S}(E^\star,Q^2,Q^2_{if}) \,,
\end{align}
where $\textbf S$ is a smooth function.

\section{Finite-volume amplitudes in 1+1D\label{sec:FV}}

Having established expressions for $\mathcal T(E^\star,Q^2 ,Q^2_{if})$ that automatically satisfy unitarity, we now turn to analogous results for the finite-volume quantity, $\mathcal T_L$, defined as 
\begin{equation}
\label{eq:TAdefFV}
\T_L(p_f, q, p_i) \equiv  2 i \sqrt{\omega_{\boldsymbol p_f} \omega_{\boldsymbol p_i}}    \, L \int d x^0 \int_0^L d x^1 \, e^{i \omega x^0 - \epsilon \vert x^0 \vert - i \bm{q} \cdot \bm{x}} \, \langle \boldsymbol p_f \vert \, \text{T} \{ \mathcal J (x) \mathcal J'(0) \} \, \vert \boldsymbol p_i \rangle_{\text{c}, L} \,,
\end{equation}
where the pre-factor is included because the single-particle states have unit normalization, whereas $2 \omega_{\bm p} L$ is the finite-volume analog of $\langle \bm p' \vert \bm p \rangle = 2 \omega_{\bm p} (2 \pi) \delta(\bm p' - \bm p)$.
Here, we treat the direction parametrized by $x_0$ to have infinite extent and to be continuous. In a quantum computation, one can only evolve with a finite number of steps in this direction, before the coherence of the calculation is lost. This presents an additional set of challenges: In particular the damping parameter, $\epsilon$, must be taken small enough to estimate the extrapolation $\epsilon \to 0$ but large enough that the integral can be estimated reliably. A more detailed discussion goes beyond the scope of this work.

\bigskip

Turning to the hadronic scattering amplitude, we begin by defining
\begin{equation}
\label{eq:MLDef}
\M_{L,[\mathcal J]}(p_f, q, p_i) =
\frac{
(Q^2 + m^2)
(Q^2_{if} + m^2)}
{\langle 0 \vert \mathcal J(0) \vert \boldsymbol q \rangle_L
\,
\langle \boldsymbol p_f+ \boldsymbol q- \boldsymbol p_i \vert \mathcal J'(0) \vert 0 \rangle_L}
\T_L(p_f, q, p_i) \,,
\end{equation}
as a rough analog of Eq.~\eqref{eq:M_from_T}. Here we have not included the on-shell limit $\{Q^2, Q^2_{if}\} \to - m^2$ as it is instructive to also consider off-shell values in understanding the extraction of the physical observable. We have also used the infinite-volume mass in the amputation. As we discuss below, using a finite-volume mass instead offers no clear advantage.

To gain better intuition, it is instructive to consider $\M_{L,[\mathcal J]}$ in the case of identical hermitian currents $\mathcal J(0) = \mathcal J'(0)$ and identical states $\boldsymbol p_f = \boldsymbol p_i = \boldsymbol p$. We evaluate the integral in Eq.~\eqref{eq:TAdefFV} by first separating the integral into $x_0>0$ and $x_0<0$ regions. We then insert a complete set of finite-volume states, to find
\begin{equation}
\M_{L,[\mathcal J]}(p, q, p)  =   - 2  \omega_{\boldsymbol p}   L^2 \frac{
(Q^2 + m^2)^2}
{ \vert \langle 0 \vert \mathcal J(0) \vert \boldsymbol q \rangle_L \vert^2} \sum_n  \frac{\vert \langle \boldsymbol p  \vert  \mathcal J(0) \vert \bm p + \bm q , n \rangle_L  \vert^2  }{  \omega  + i \epsilon    + \omega_{\bm p}  - E_n^{\bm p + \bm q}(L)  } + \big [ (\omega, \boldsymbol q) \to -(  \omega,   \boldsymbol q)  \big ] - \cdots \,,
\label{eq:MLDecom2}
\end{equation}
where $\vert \boldsymbol p+ \boldsymbol q, n \rangle_L$ is the $n$th finite-volume excited state with the indicated momentum and $E^{\bm p + \bm q}_n(L)$ is its corresponding energy. Here the $(\omega, \boldsymbol q) \to -(  \omega,   \boldsymbol q)$ contribution arises from the other time ordering while the ellipsis indicates disconnected contributions. These must be treated separately when inserting a complete set of states, but they play no role in the discussion here.

In the $L \to \infty$ limit, the sum over discrete states goes over to an integral and the set $\big \{ \vert \boldsymbol p+ \boldsymbol q, n \rangle_L \big \}$ goes over to a continuum of multi-particle states (either in or out states, both choices are viable). Then the corresponding matrix elements are labeled not only by total four-momentum, but also the momentum of the incoming particles, e.g.~$\langle \boldsymbol p \vert \mathcal J(0) \vert q + p, \boldsymbol p', \text{in} \rangle$ for a two-particle state. Such matrix elements contain a disconnected term, proportional to $\delta^3(\boldsymbol p - \boldsymbol p')$, which, in the integral over individual particle momenta, generates the factors of $ \langle 0 \vert \mathcal J(0) \vert \boldsymbol q \rangle / (Q^2 + m^2)$. The amputation of these then yields the infinite-volume hadronic amplitude, $\mathcal M(E^\star)$.

By contrast, at finite $L$, $\mathcal M_{L, [\mathcal J]}$ exhibits a discrete set of poles at $\omega = \pm [E^{\bm p \pm \bm q}_n(L) - \omega_{\boldsymbol p} - i \epsilon]$, none of which directly correspond to the $\sim 1/(Q^2 + m^2)$ factors we are after. As a result
\begin{equation}
\label{eq:ZeroOnShell}
\lim_{Q^2 \, \to \, - m^2} \M_{L,[\mathcal J]}(p, q, p) = 0 \,,
\end{equation}
so that the naive on-shell limit offers no useful information about the target amplitude. The same result holds for the finite-volume mass, provided the amputation factor is adjusted as well. Indeed since the finite-volume quantity only has simple poles, the $[Q^2 + m^2]^2$ amputation will always lead to a vanishing on-shell limit, unless one defines the limit such that the amputating factor remains non-zero. As explained in Ref.~\cite{Bulava:2019kbi}, this observation fits naturally with the fact that we send $L \to \infty$ before $\epsilon \to 0$; the non-zero epsilon provides a non-zero amputation factor. A related issue is that the matrix element, $\langle \boldsymbol p \vert \mathcal J (0) \vert \boldsymbol p+ \boldsymbol q, n \rangle_L$, does not factorize into a vacuum-to-single-particle component at finite-$L$, so the cancellation of $\mathcal J(0)$-dependence in Eq.~\eqref{eq:MLDecom2} is obscured.

Despite these complications, for arbitrarily large $L$ a correspondence to the infinite-volume quantities must be recovered. As we now describe, this can be understood in detail using the finite-volume formalism presented in Ref.~\cite{Briceno:2019opb}. To explain this, we focus first on the finite-volume Compton analog, $\T_L(p_f, q, p_i)$, for which $\mathcal J(0)$-dependence is a natural feature of the amplitude's definition. An analytic expression for the $L$ dependence of this quantity was derived in \cite{Briceno:2019opb}, for a generic 3+1D scalar field theory. The simplification of this to 1+1D only requires modifying the definition of a finite-volume geometric function, denoted $F$, as we explain below. We can thus take over the main result directly 
\begin{align}
\label{eq:TL}
\T_{L}(p_f,q,p_i) =
\T (E^\star,Q^2 ,Q^2_{if}) -
\cH(E^\star ,Q^2)
\,
\frac{1}{F^{-1}(E^\star,\boldsymbol P, L) + \M(E^\star) } 
\,
\cH'(E^\star ,Q^2_{if})
\,.
\end{align}
This is the most important result of this section and will serve as our master formula for the numerical analysis presented in Sec.~\ref{sec:numerical}. The result is exact up to terms scaling as $e^{- m L}$ and holds for any relativistic quantum field theory. All infinite-volume amplitudes appearing here are defined in the previous section, and the only new object is $F(E,\boldsymbol P, L)$. 

A diagrammatic representation of $\T_L$ is shown in Fig.~\ref{fig:iTL}($a$). As indicated by the figure, $\T_L$ is defined via the same expansion as its infinite-volume analog, $\T$, but with all internal loops summed, rather than integrated, over the spatial momenta consistent with the finite-volume boundary conditions. For a periodic volume, the allowed set is given by $\boldsymbol k = 2 \pi \boldsymbol n/L$ with $\boldsymbol n \in \mathbb Z$ an integer. The new quantity, $F(E,\boldsymbol P, L)$, is a geometric function that contains a sum-integral difference, encoding the distinction between finite- and infinite-volume amplitudes. The explicit expression is
\begin{align}
F(E,\boldsymbol P, L)
&= \lim_{\epsilon \to 0^+}
\frac12\left[\frac{1}{L}\sum_{\boldsymbol k}-\int\frac{d \boldsymbol k}{2\pi}\right]\frac{1}{2\omega_k} \frac{1}{(P-k)^2-m^2+i\epsilon},
\label{eq:Fdef}
\\
&= i \rho(E^\star)
+ 
\frac{\rho(E^\star)}{2}
\left [
\cot \left(\frac{L \gamma (q^\star+\omega_q^\star \beta) }{2}\right)
+
\cot \left( \frac{L \gamma (q^\star-\omega_q^\star \beta) }{2}\right)
\right ] + \mathcal O(e^{- m L})
\,,
\label{eq:Fcot}
\end{align}
where $\omega_q^\star=\sqrt{q^{\star 2}+m^2} = E^\star/2$, $\gamma = E/E^\star$ and $\beta = \boldsymbol P /E$. The second line here is derived in Appendix~\ref{app:ps}. In the following equations we will drop the neglected $ \mathcal O(e^{- m L})$ corrections.

As discussed at the beginning of this section, $\T_L$ has poles on the real axis, depicted in Fig.~\ref{fig:iTL}($b$). The poles correspond to the finite-volume energy levels and therefore provide the 1+1D analog of the L\"uscher quantization condition
\begin{align}
{\M(E^\star)^{-1} + F(E,\boldsymbol P, L) } 
&= 0 \,.
\label{eq:QC}
\end{align}
Combining Eqs.~\eqref{eq:Mdef}, \eqref{eq:cotd}, and \eqref{eq:Fcot}, we can rewrite this condition as
\begin{align}
{\cot\delta(E^\star) +\frac{1}{2}
\bigg [
\cot \left(\frac{L \gamma (q^\star+\omega_q^\star \beta) }{2}\right)
+
\cot \left( \frac{L \gamma (q^\star-\omega_q^\star \beta) }{2}\right)
\bigg ]
}=0 \,.
\label{eq:QCv2}
\end{align}
For the special case of $\boldsymbol P=0$ we recover the well-known result, that finite-volume energies satisfy \cite{\OneDimQC}
\begin{align}
q^\star_n(L) L = - 2 \delta(E^\star_n(L)) + 2 \pi n \,, \ \ \ \ \ \ \ \ E^\star_n(L) = 2 \sqrt{m^2 + q^\star_n(L)^2} \,, \ \ \ \ \ \ \ \ E_n(L) = \sqrt{E^\star_n(L)^2 + \boldsymbol P^2}\,.
\end{align}

As written, however, Eq.~\eqref{eq:TL} does not encode the non-infinitisimal $\epsilon$ dependence appearing in the definition of $\T_L$, Eq.~\eqref{eq:TAdefFV}. The $\epsilon$ dependence can be incorporated, to good approximation, by replacing $E$ with $E+ i \epsilon$ in all kinematic expressions, while keeping $\boldsymbol P$ independent of this parameter. The only subtlety here is that, as can be seen in Eq.~\eqref{eq:MLDecom2}, both $E + i \epsilon$ and $E - i \epsilon$ enter the original definition of $\T_L$. However, in the decomposition leading to \eqref{eq:TL} only the former combination ($E + i \epsilon$) enters in the power-like finite-volume effects that we keep. The anti-particle prescription ($E - i \epsilon$) is absorbed into the infinite-volume quantities. Thus setting $E \to E+ i \epsilon$ everywhere amounts to neglecting terms scaling as $\epsilon/\mu$, where $\mu$ is the smallest scale entering the infinite-volume amplitudes.

\bigskip

To summarize the results so far, our master equation, Eq.~\eqref{eq:TL}, gives an expression for the $L$ dependence of $\T_L$, in terms of expressions for the infinite-volume amplitudes $\T$, $\mathcal H$, $\mathcal H'$ and $\mathcal M$. This provides a tool to explore optimal numerical strategies for approaching the physical amplitude in future Minkowski-signature calculations, especially for non-perturbative systems. These numerical explorations are the focus on the next section. 

Before turning to this, we consider a handful of formal results that follow directly from Eq.~\eqref{eq:TL}. As discussed above, and also in Refs.~\cite{Hansen:2015zga, Hansen:2017mnd, Bulava:2019kbi,\OpticalMai} only the ordered double limit, $L \to \infty$ followed by $\epsilon \to 0$, is expected to recover the physical amplitude. Our general expression, Eq.~\eqref{eq:TL}, reproduces this fact trivially via
\begin{align}
\lim_{\epsilon\to 0}\lim_{L\to \infty}F(E +i\epsilon,\boldsymbol P, L)&=
\lim_{\epsilon\to 0}\lim_{L\to \infty}\frac12 \left[\frac{1}{L}\sum_{k}-\int\frac{dk}{2\pi}\right] \frac{1}{2\omega_k} \frac{1}{(E-\omega_k + i\epsilon)^2- (\boldsymbol P - \boldsymbol k)^2 - m^2},
\nn\\
&=
\lim_{\epsilon\to 0}\frac12 \left[\int\frac{dk}{2\pi}-\int\frac{dk}{2\pi}\right] \frac{1}{2\omega_k} \frac{1}{(E-\omega_k + i\epsilon)^2- (\boldsymbol P - \boldsymbol k)^2 - m^2},
\nn\\
&=
0 \,.
\label{eq:F_lim}
\end{align}
Applying this to Eq.~\eqref{eq:TL} then directly implies
\begin{align}
\lim_{\epsilon\to 0}\lim_{L\to \infty}
\T_{L}(p_f,q,p_i) 
=
\T (E^\star,Q^2 ,Q^2_{if}) \,.
\label{eq:IV_lim}
\end{align}
Note also that, although the finite-volume amplitude depends on six variables, via the three two-component vectors ($p_f,q,p_i$), its infinite-volume counterpart only depends on three Lorentz scalars, $E^\star=\sqrt{P^2}$, $Q^2$, and $Q^2_{if}$. We make use of this observation in Sec.~\ref{sec:numerical}.

To understand the approach towards infinite $L$ it is also instructive to note that, if $\epsilon$ and $L$ are chosen such that $\mathcal M(E^\star) F(E + i \epsilon, \boldsymbol P, L) \ll 1$, then we can expand Eq.~\eqref{eq:TL} as
\begin{equation}
\T_{L}(p_f,q,p_i) =
\T (E^\star,Q^2 ,Q^2_{if})
-
\cH'(E^\star ,Q^2)
\,
F(E + i \epsilon,\boldsymbol P, L) 
\,
\cH (E^\star ,Q^2_{if})
+ \mathcal O(F^2)
\,.
\end{equation}
This will motivate one of the strategies we consider in the next section in which we identify sets of kinematics that leave $\T (E^\star,Q^2 ,Q^2_{if})$ invariant and show that averaging over these suppresses $F(E + i \epsilon,\boldsymbol P, L) $ and therefore improves the infinite-volume extrapolation.

To close this section we return to the hadronic amplitude, $\mathcal M(E^\star)$, and the finite-volume quantity defined in Eq.~\eqref{eq:MLDef}. Combining this with our master equation \eqref{eq:TL} we reach
\begin{align}
\label{eq:MLfromTL}
\M_{L,[\mathcal J]}(p_f, q, p_i) =
\M_{[\mathcal J]}(E^\star,Q^2 ,Q^2_{if}) 
-
\M_{[\mathcal J]}(E^\star ,Q^2)
\,
\frac{1}{F^{-1}(E^\star,\boldsymbol P, L) + \M(E^\star) } 
\,
\M'_{[\mathcal J]}(E^\star ,Q^2_{if}) 
\,,
\end{align}
where
\begin{align}
\label{eq:MDOffDef}
\M_{[\mathcal J]}(E^\star,Q^2 ,Q^2_{if}) & \equiv \frac{
(Q^2 + m^2)
(Q^2_{if} + m^2)}
{\langle 0 \vert \mathcal J(0) \vert \boldsymbol q \rangle
\,
\langle \boldsymbol p_f+ \boldsymbol q- \boldsymbol p_i \vert \mathcal J'(0) \vert 0 \rangle}
\T (E^\star,Q^2 ,Q^2_{if}) \,, \\
\M_{[\mathcal J]}(E^\star ,Q^2) & \equiv \frac{
(Q^2 + m^2)}
{\langle 0 \vert \mathcal J(0) \vert \boldsymbol q \rangle}
\cH (E^\star,Q^2) \,, \\
\M'_{[\mathcal J]}(E^\star ,Q^2_{if}) & \equiv \frac{
(Q^2_{if} + m^2)}
{ \langle \boldsymbol p_f+ \boldsymbol q- \boldsymbol p_i \vert \mathcal J'(0) \vert 0 \rangle}
\cH'(E^\star ,Q^2_{if}) \,.
\label{eq:MpOffDef}
\end{align}
The notation here emphasizes that the quantities in Eqs.~\eqref{eq:MDOffDef}-\eqref{eq:MpOffDef} depend on at least one virtuality, $\{Q^2 ,Q^2_{if}\}$, as well as the details of the currents $\mathcal J(0)$, $\mathcal J'(0)$. In the on-shell limit however, this dependence is removed and each object corresponds with the on-shell hadronic amplitude, e.g.
\begin{equation}
\lim_{Q^2, Q^2_{if} \to -m^2} \M_{[\mathcal J]}(E^\star,Q^2 ,Q^2_{if}) = \mathcal M(E^\star) \,.
\end{equation}
This, together with Eq.~\eqref{eq:MLfromTL}, then implies
\begin{equation}
\label{eq:MLJExpQ}
\M_{L,[\mathcal J]}(p_f, q, p_i) = \mathcal M_L(E, \boldsymbol P) + \mathcal O\big [ (Q^2 + m^2), (Q^2_{if} + m^2)\big ] \,,
\end{equation}
where we have introduced
\begin{equation}
\label{eq:MLdefNoJ}
\mathcal M_L(E, \boldsymbol P) \equiv \frac{1}{\mathcal M(E^\star)^{-1} + F(E, \boldsymbol P, L) } \,.
\end{equation}
The function $\mathcal M_L(E, \boldsymbol P)$ arises often in the context of finite-volume quantization conditions, see for example Refs.~\cite{\MLlit}.

A slightly confusing point is the consistency of these equations with Eq.~\eqref{eq:ZeroOnShell} above, i.e.~the observation that $\M_{L,[\mathcal J]}(p_f, q, p_i)$ vanishes in the on-shell limit. To see that the results are consistent note first that, in the forward case ($\boldsymbol p_f = \boldsymbol p_i$), the on-shell condition is achieved by setting $\omega = \omega_{\boldsymbol q}$, equivalently setting $E = \omega_{\boldsymbol p} + \omega_{\boldsymbol q}$. At such energies $F$ diverges
\begin{equation}
\lim_{E \, \to \, \omega_{\boldsymbol p} + \omega_{\boldsymbol q}} F(E ,\boldsymbol P, L) = \infty \,,
\end{equation}
as can be easily seen from Eq.~\eqref{eq:Fcot}, and thus
\begin{equation}
\lim_{E \, \to \,  \omega_{\boldsymbol p} + \omega_{\boldsymbol q}} \mathcal M_L(E, \boldsymbol P) = 0 \,.
\end{equation}
Given that $\mathcal M_L(E, \boldsymbol P)$ and $\M_{L,[\mathcal J]}(p_f, q, p_i) $ only coincide for kinematics where they are also identically zero, one might question how any of these expressions can be useful. Again the resolution is the $i \epsilon$ prescription. Repeating the steps above with nonzero $i \epsilon$ gives
\begin{equation}
\M_{L,[\mathcal J]}(p_f, q, p_i) \bigg \vert_{E \to E + i \epsilon} = \mathcal M_L(E+ i \epsilon, \boldsymbol P) + \mathcal O\big [ (Q^2 + m^2), (Q^2_{if} + m^2)\big ] \,,
\end{equation}
where, in the final term, $q = (\omega + i \epsilon, \boldsymbol q)$. Now if we set $E =  \omega_{\boldsymbol p} + \omega_{\boldsymbol q}$ then we recover a non-zero value of $\mathcal M_L(E+ i \epsilon, \boldsymbol P)$ which estimates the amplitude up to corrections of order $(Q^2 + m^2) = \mathcal O(\epsilon)$.

\section{Ordered double limit: Challenges and strategies\label{sec:numerical}}

In this section, we discuss strategies for numerically recovering the infinite-volume amplitudes, $\mathcal M(E^\star)$ and $\T(E^\star,Q^2 ,Q^2_{if})$, from their finite-volume counterparts. To do so, we require plausible functional forms for the infinite-volume quantities entering our master equation, Eq.~\eqref{eq:TL}. As described in the previous two sections, $\T_L(p_f,q,p_i)$ is ultimately given by four real functions, $\K(E^\star)$, $\bH(E^\star,Q^2)$, $\bH'(E^\star,Q^2)$ and $\mathbf{S}(E^\star,Q^2,Q^2_{if})$. Beginning with the K matrix, we write 
\begin{gather}
\K (E^{\star}) =  m^2 q^{\star 2} \bigg(\frac{g^2}{m_{R}^2 - E^{\star2}}+h(E^{\star2}) \bigg) \,,
\label{eq:Kmatpar} 
\end{gather}
where $g$ is a dimensionless coupling (such that $\mathcal K(E^\star)$ has dimensions of $m^2$), $m_R$ is an independent parameter with units of energy, and $h(E^{\star2})$ is a polynomial in $E^{\star2}$ (also with dimension $m^2$).

Two basic assumptions motivate this parametrization of $\mathcal K(E^\star)$. First, we require that the system has no threshold singularities. This leads to the overall factor of $q^{\star 2}$ which regulates the near threshold behavior of the 1+1D amplitude. Second, we assume that left-hand cuts, inelastic thresholds, and other analytic structures within $\mathcal K(E^\star)$ are sufficiently removed from the energy region sampled, such that their effects can be well described by the given form. The free parameters entering Eq.~\eqref{eq:Kmatpar} afford a great deal of freedom in the systems that can be described. One can choose the parameters to describe weakly or strongly interacting systems, including systems with a broad or narrow resonance, as well as a bound state.

For $\bH$, $\bH'$ and $\textbf S$, we use minimal expressions that satisfy the criteria discussed in Sec.~\ref{sec:IV_amps}, 
\begin{equation}
\label{eq:Hparam}
\bH(E^\star, Q^2) = \bH'(E^\star, Q^2) = 
\frac{\K(E^\star)}{1+{Q^2}/M^2} \,, \qquad \qquad
\mathbf{S}(E^\star,Q^2,Q_{if}^2) = 0 \,.
\end{equation}
Here we have enforced that $\bH(E^\star, Q^2)$ must have the same poles as $\K(E^\star)$. We have additionally introduced a time-like pole in the $Q^2$ dependence of $\bH(E^\star, Q^2)$ and $\bH'(E^\star, Q^2)$ to mimic the known behavior of certain form factors.

Before turning to the numerical studies, we revisit Eq.~\eqref{eq:MLDef} in which an estimator for the hadronic amplitude (denoted $\M_{L,[\mathcal J]}$) is constructed from $\T_L$. Setting $p_f = p_i = p$, restricting attention to $\textbf S = 0$, and rearranging the result to express $\M_{L,[\mathcal J]}$ directly in terms of $\bH$ and $\mathcal K$, we reach
\begin{equation}
\M_{L,[\mathcal J]}(p, q, p) = \frac{
(Q^2 + m^2)^2 \,\, \bH(E^\star,Q^2)^2}
{\vert \langle 0 \vert \mathcal J(0) \vert \boldsymbol q \rangle_L \vert^2} 
\bigg [
\frac{1}{\K(E^\star)} -
\frac{1}{1 + F_{\text{pv}}(E, \boldsymbol P, L) \K(E^\star) } F_{\text{pv}}(E, \boldsymbol P, L) 
\bigg ]\,,
\end{equation}
where we have defined $F_{\text{pv}}(E, \boldsymbol P, L) \equiv F(E, \boldsymbol P, L) - i \rho(E^\star) $ and followed standard algebraic manipulations to remove $\mathcal H$ and $\mathcal M$ in favor of $\bH$ and $\mathcal K$.

In the case where $\mathcal J$ has the quantum numbers of the single-hadron interpolator (as required for $\M_{L,[\mathcal J]}$), the function $\bH(E^\star, Q^2)$ must contain a pole at $Q^2 = m^2$. This can easily be encoded in our more general parametrization by setting $M = m$. Additionally using the fact that the current must satisfy $\vert \langle 0 \vert \mathcal J(0) \vert \boldsymbol q \rangle_L \vert^2 = m^4$, we finally reach\footnote{To see why $\vert \langle 0 \vert \mathcal J(0) \vert \boldsymbol q \rangle_L \vert^2 = m^4$, note first that the single-particle matrix element has only exponentially-suppressed finite-volume effects, neglected throughout, and must thus equal some combination of physical parameters defining the infinite-volume theory. In addition, the LSZ reduction formula demands that $\M_{L,[\mathcal J]}$ will become the physical scattering amplitude in the ordered double limit. These constraints are enough to give the claimed result.}
\begin{align}
\M_{L,[\mathcal J]}(p, q, p) & = \mathcal K(E^\star)^2
\bigg [
\frac{1}{\K(E^\star)} -
\frac{1}{1 + F_{\text{pv}}(E, \boldsymbol P, L) \K(E^\star) } F_{\text{pv}}(E, \boldsymbol P, L) 
\bigg ] \,, \\
& = \frac{1}{\mathcal K(E^\star)^{-1} + F_{\text{pv}}(E, \boldsymbol P, L) } \,, \\
& = \frac{1}{\mathcal M(E^\star)^{-1} + F(E, \boldsymbol P, L) } \,, \\[7pt]
& = \mathcal M_L(E, \boldsymbol P) \,.
\label{eq:MLqc}
\end{align}
This result closely resembles Eq.~\eqref{eq:MLJExpQ} but without the $Q^2+m^2$ corrections. The interpretation is that, for our specific model of $\bH, \bH'$, and for $\textbf S = 0$, the higher corrections vanish. Though unrealistic in a practical calculation, this set-up provides a useful starting point in our analysis of finite-volume contaminations.

\subsection{Hadronic amplitude: Basic estimators\label{sec:fixedvolHad}}

\begin{figure}[t!]
\begin{center}
\includegraphics[width=1\textwidth]{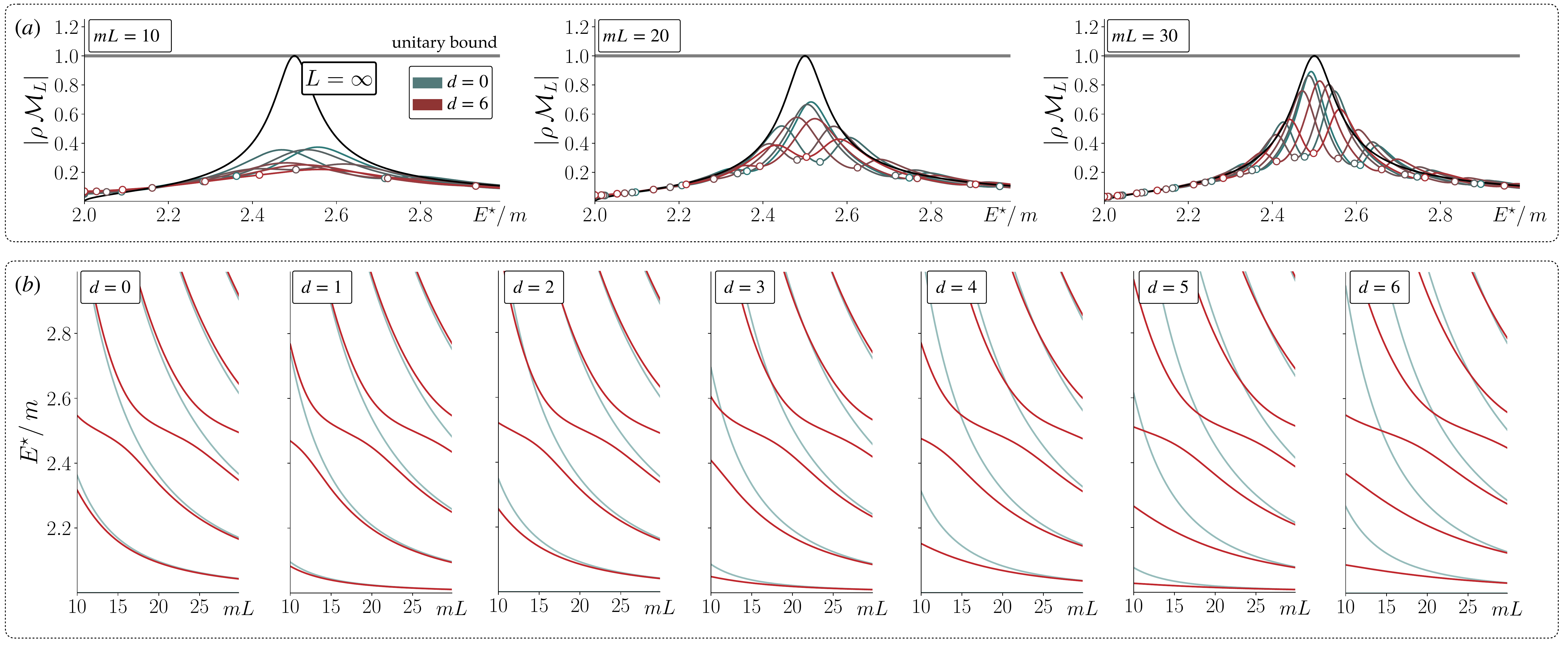}
\caption{$(a)$ Finite- and infinite-volume amplitudes (multi-color and black curves respectively), evaluated using $m_R=2.5m$, $g=2.5$, $h(E^{\star2}) = 0$ as described in the text. For the finite-volume amplitude, defined in Eq.~\eqref{eq:MLdefNoJ}, we consider seven different momenta ($\boldsymbol d = 0,1,\dots,6$, with $\boldsymbol P = 2 \pi \boldsymbol d/L$) and set $\epsilon =1/L$ in each panel. As explained after Eq.~\eqref{eq:unitary_bound}, the small circles indicate the points for which $Q^2 + m^2 = \mathcal O(\epsilon)$. $(b)$ Finite-volume spectrum for the non-interacting system (light-blue) and interacting system (red). \label{fig:boosting}} 
\end{center}
\end{figure}

We begin by numerically exploring the convergence of $\mathcal M_L(E+ i \epsilon, \boldsymbol P)$ towards the infinite-volume hadronic amplitude, $\mathcal M(E^\star)$. We focus on the case of a resonance, with parameters $m_R=2.5m$ and $g=2.5$, and with $h(E^{\star2}) = 0$. With this parametrization, in Fig.~\ref{fig:boosting}($a$) we compare $\mathcal M_L(E + i \epsilon, \boldsymbol P)$ and $\mathcal M(E^\star)$, both plotted versus $E^\star$, for various fixed values of $L$, $\epsilon$ and $\boldsymbol P$. Specifically we plot the combinations $\vert \rho(E^\star) \mathcal M_L(E + i \epsilon, \boldsymbol P) \vert $ and $\vert \rho(E^\star) \mathcal M(E^\star) \vert $, which make the unitarity bound particularly transparent 
\begin{align}
\vert \rho(E^\star) \mathcal M(E^\star) \vert = \left|\frac{1}{\cot\delta(E^\star)-i}\right|\leq 1 \,.
\label{eq:unitary_bound}
\end{align}
The three panels correspond to three spatial volumes, each displaying seven curves corresponding to spatial momenta ranging from $\boldsymbol d=0$ to $\boldsymbol d = 6$, with $\boldsymbol P = 2 \pi \boldsymbol d/L$. 
The circles in Fig.~\ref{fig:boosting}($a$) indicate the finite-volume energies for which $Q^2 + m^2 = \mathcal O(\epsilon)$ can be achieved. 

The energies for which this matching is possible are also the non-interacting levels of the system, given by
\begin{equation}
E(L) = \sqrt{m^2 + (2 \pi/L)^2 \boldsymbol n^2} + \sqrt{m^2 + (2 \pi/L)^2 (\boldsymbol n- \boldsymbol d)^2} \,,
\end{equation}
and represented in Fig.~\ref{fig:boosting}($b$) as the blue curves. The functional form of $\mathcal M_L(E + i \epsilon, \boldsymbol P)$, by contrast, is dictated by the interacting spectrum, obtained using Eq.~\eqref{eq:QCv2} and plotted in Fig.~\ref{fig:boosting}($b$) as the red curves.

\begin{figure}[t!]
\begin{center}
\includegraphics[width=1\textwidth]{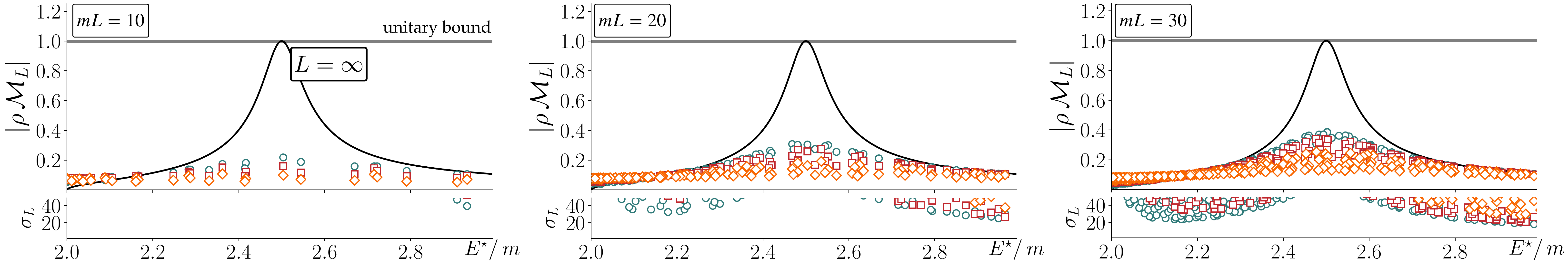}
\caption{As in Fig.~\ref{fig:boosting}($a$), but here we only show the points for which $Q^2 + m^2 = O(\epsilon)$, as described in the text. The set of total momenta also matches Fig.~\ref{fig:boosting}($a$), but is collected here as single color. The blue-green circles, red squares, and orange diamonds correspond to $\epsilon L=1$, $\epsilon L=2$ and $\epsilon L=4$ respectively.  The small lower plots in each panel show $\sigma_L$, defined in Eq.~\eqref{eq:sigmaL}, which measures the percent deviation from the physical scattering amplitude. For many points this exceeds $50\%$ and is above the plotted range.\label{fig:iML_disc}
 }
\end{center}
\end{figure}

In Fig.~\ref{fig:iML_disc}, we show $\M_L(E+i \epsilon, \boldsymbol P)$ for the $Q^2 + m^2 = \mathcal O(\epsilon)$ points, taking all $\boldsymbol P$ values together and considering three possible values of $\epsilon L=4,2,1$. From this figure it is quite evident that the convergence to the infinite-volume amplitude is slow. To quantify the deviation, we introduce
\begin{align}
\sigma_L(E^\star, \boldsymbol P, \epsilon)=100\times\left|
\frac{\M_L(E + i \epsilon, \boldsymbol P)-\M(E^\star)}{\M(E^\star)}
\right| \,,
\label{eq:sigmaL}
\end{align}
plotted in the small lower panels of Fig.~\ref{fig:iML_disc}. Even for $mL=30$ the systematic uncertainty is well above $20 \%$ for a wide range of energies. Further investigation finds that volumes in the order of $mL=10^2-10^3$ are required to recover amplitudes at the percent level from this approach.

These results call for improved strategies in extracting the amplitude. For example, note that the curves in Fig.~\ref{fig:boosting}($a$) oscillate as a function of $E^\star$, about some underlying curve, for any given values of $\epsilon$ and $L$. Smaller $\epsilon$ values lead to oscillations with higher amplitude and lower frequency, and the relative phases depend also on the specific choice of $\boldsymbol P$. This suggests that the average over curves defined with different $\boldsymbol P$ may approach $L \to \infty$ more quickly than the individual functions. We explore this strategy in the context of the Compton amplitude, in the next subsection.

The Compton amplitude, $\mathcal T$, proves more instructive than the hadronic amplitude, $\mathcal M$, for two reasons. First, $\mathcal M_L$ is only an approximation to the quantity, $\mathcal M_{L,[\mathcal J]}$, that is in principle accessible from a Minkowski correlator. In particular, the $\mathcal J(0)$ dependence of $\mathcal M_{L,[\mathcal J]}$ vanishes only in the infinite-volume limit and this determines how the limit is approached. Modeling such operator dependence goes beyond the scope of this work. By contrast, for the Compton amplitude, the dependence on the details of $\mathcal J(0)$ is a natural feature that persists in the final observable. Second, in the 1+1D theory, the Compton amplitude depends on three Lorentz invariants, in contrast to the dependence on $s$ that arises with $\mathcal M$. This presents additional challenges and opportunities, as we now describe.

\subsection{Compton amplitude: Binning over similar kinematics\label{sec:bin}}

We continue to use the parameterizations described in Eqs.~\eqref{eq:Kmatpar} and \eqref{eq:Hparam} with the K-matrix parameters of the previous section ($m_R=2.5m$, $g=2.5$, $h(E^{\star2}) = 0$) but now setting $M = m_R$ for the form-factor mass within $\bH$. As $\mathcal T(E^\star, Q^2, Q_{if}^2)$ has more degrees of freedom than $\mathcal M(E^\star)$, one requires an approach to estimate and present the more complicated functional form. We find it most instructive to plot slices of the Compton amplitude, defined with certain kinematics fixed. To achieve this we define the following function
\begin{equation}
{\overline {\mathcal T_L}}\big (\overline {E^\star}, \overline {Q^2} \big) = \frac{1}{\mathcal N} \sum_{L, \epsilon} \sum_{\{ \boldsymbol q, \boldsymbol p_f, \boldsymbol p_i, \omega \} \in \Omega} \delta \big (\boldsymbol q, \boldsymbol p_f, \boldsymbol p_i, \omega \vert \overline {E^\star}, \overline {Q^2} \big ) \, \mathcal T_{L}(p_f,q,p_i) \,,
\label{eq:bin}
\end{equation}
where $\delta = 0$ or $1$ based on a binning criteria and $\mathcal N$ counts the number of contributions, to normalize the average in a given bin. We have also introduced $\Omega$ to represent the set of all redundant kinematics over which the average may be performed. Finally the left-hand side only depends on a single virtuality $\overline {Q^2}$ as we enforce $Q^2 = Q^2_{if}$ in all plots considered in this section.

\begin{figure}[t!]
\begin{center}
\includegraphics[width=1\textwidth]{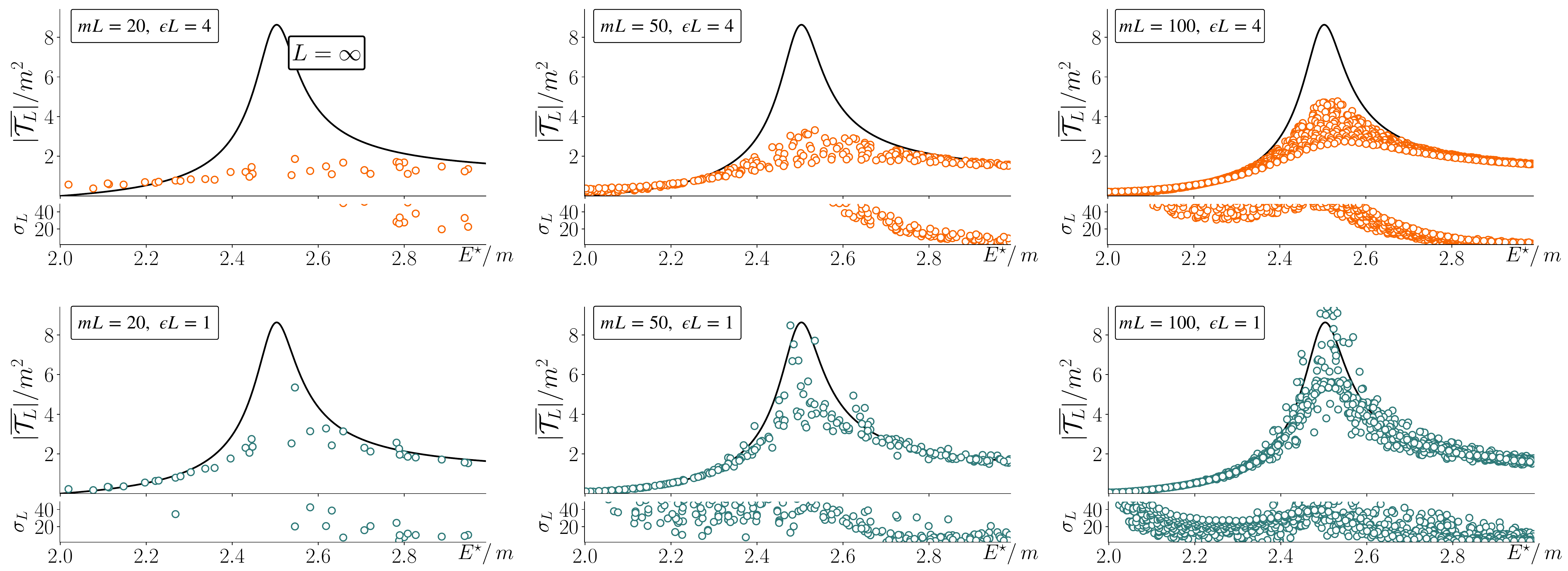}
\caption{Comparison of the infinite-volume amplitude, $\T$ (black curve), with the finite-volume estimator $\overline {\T_L}$, defined in Eq.~\eqref{eq:bin}, (colored points) both with $Q^2=Q^2_{if}=2 m^2$. Here we use the same K-matrix parametrization as in Fig.~\ref{fig:iML_disc} [see also Eqs.~\eqref{eq:Kmatpar} and \eqref{eq:Hparam}]. The details of the binning are given in Eq.~\eqref{eq:BinSet1} and the surrounding text. Finally, the small lower plots on each panel indicate the percent deviation, with $\sigma_L$ defined in Eq.~\eqref{eq:sigmaL}, with $\mathcal T$ in place of $\mathcal M$. \label{fig:iTL_disc}}
\end{center}
\end{figure}

As a first example, in Fig.~\ref{fig:iTL_disc} we consider fixed values of $L$ and $\epsilon$ ($mL=20,50,100$ and $\epsilon L=1,4$, as indicated in the plot), and also perform no averaging over energy, i.e.~we set $E^\star = \overline {E^\star}$. The binning procedure thus runs only over $Q^2$ and $Q_{if}^2$ and is defined by taking $\delta = 1$ whenever 
\begin{equation}
\label{eq:BinSet1}
\big |\overline{Q^2}-{Q}^2 \big | < \Delta_{Q^2} \ \ \ \ \ \text{and} \ \ \ \ \ \big | {Q}_{if}^2-{Q}^2 \big | < \Delta_{Q^2} \,,
\end{equation}
and $\delta = 0$ otherwise. Here we fix the target value to $\overline{Q^2} = 2m^2$ and the resolution to $\Delta_{Q^2}=0.01 m^2$. In the set $\Omega$ we include all $\boldsymbol{p}_f$, $\boldsymbol{p}_i$, and $\boldsymbol{q}$ sampled independently from $-2\pi \boldsymbol d_{\rm max}/L$ to $2\pi \boldsymbol d_{\rm max}/L$ in discrete steps of $2\pi/L$, where we take $\boldsymbol d_{\rm max}=mL$. The latter is a somewhat arbitrary choice motivated by the fact that the number of available modes scales with $L$ if the spatial discretization is held fixed. As $\omega$ is continuous, we simply fix the value such that $\overline{Q^2}-{Q}^2 = 0$ in the $\epsilon \to 0$ limit. The result of this construction is that $\overline{\mathcal T_L}$ continues to exhibit dramatic deviations from the infinite-volume amplitude, especially in the resonance region. This is also emphasized in the bottom panel $\sigma_L$ plots, where the definition for $\mathcal T$ is inherited from that for $\mathcal M$, Eq.~\eqref{eq:sigmaL}.

\begin{figure}[t!]
\begin{center}
\includegraphics[width=1\textwidth]{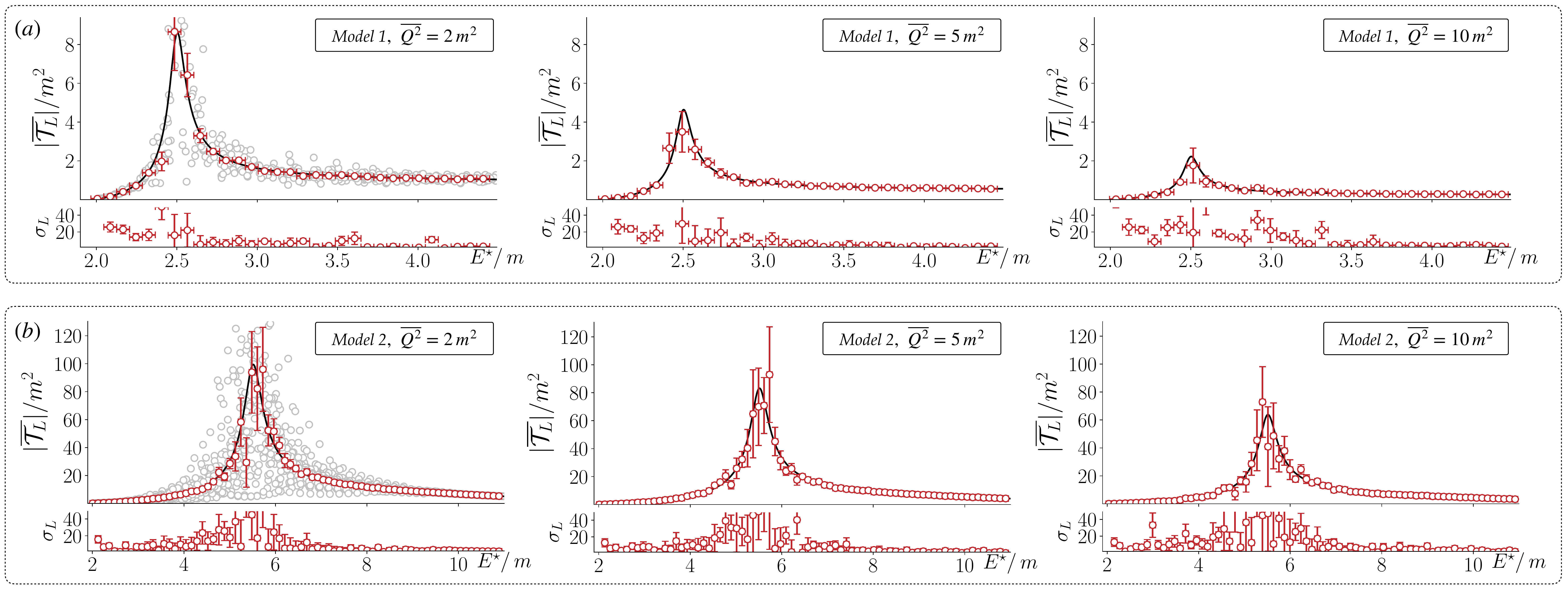}
\caption{ 
$(a)$ Comparison of the infinite-volume amplitude, $\T$ (black curve), with the finite-volume estimator detailed in Eqs.~\eqref{eq:bin} and in the paragraph containing Eq.~\eqref{eq:BinSet2} (red points). Here all data is generated according to the \emph{Model 1} parameter set (also used in Figs.~\ref{fig:boosting}, \ref{fig:iML_disc}, and \ref{fig:iTL_disc}): $m_R=2.5m$, $g=2.5$, $h(E^{\star2}) = 0$. $(b)$ As with $(a)$ but using the \emph{Model 2} parameter set: $m_R=5.5m$, $g=6$, $h(E^{\star2}) = 0.2/m^2$. The light grey points in the two leftmost panels show the set of underlying values for $\mathcal T_L$ that enter the bins defining $\overline{\mathcal T_L}$.
\label{fig:iTL_binning}}
\end{center}
\end{figure}

To improve the situation further, in Fig.~\ref{fig:iTL_binning} we average over $mL=20,25,30$, to cancel fluctuations associated with a specific $L$ value, and we further include $E^\star$ as a binned coordinate within $\Omega$. For the latter we sample $\overline {\mathcal T_L}$ in discrete steps of $\overline {E^\star}$, separated by $2 \Delta_{E^\star}$ where $\Delta_{E^\star} = 0.08m$. We then evaluate Eq.~\eqref{eq:bin} with $\delta = 1$ whenever
\begin{equation}
\big |\overline {Q^2}-{Q}^2 \big | < \Delta_{Q^2} \ \ \ \ \ \text{and} \ \ \ \ \ \big | {Q}_{if}^2-{Q}^2 \big | < \Delta_{Q^2} \ \ \ \ \ \text{and} \ \ \ \ \ |\overline {E^\star} - E^\star| \leq \Delta_{E^\star} \,.
\label{eq:BinSet2}
\end{equation}
Finally we use a more aggressive choice in $\epsilon$ as compared to the previous plot and a slightly coarser binning in virtualities: $\epsilon(L)=1/(L (mL)^{1/2})$ and $\Delta_{Q^2}=0.05\,m^2$. 

The procedure defining Fig.~\ref{fig:iTL_binning}, the most complicated considered in this work, also achieves the best reconstruction of the infinite-volume amplitude. To check whether this is robust we apply the procedure to six different functions, defined by three choices of virtualities [$\overline{Q^2} = 2m^2, 5m^2, 10m^2$] together with two different models for the underlying resonance parameters. For the latter we define \emph{Model 1} with $m_R = 2.5m$, $g=2.5$, $h(E^{\star2}) = 0$, as above, followed by \emph{Model 2}, with $m_R=5.5m$, $g=6$, $h(E^{\star2}) = 0.2/m^2$; see again Eq.~\eqref{eq:Kmatpar}. Note that the reconstruction is less effective in reproducing singularities of the amplitude. This is evident from the spikes in $\sigma_L$ close to the threshold (a kinematic singularity) and near the resonant peak (near the dynamical singularity of the resonance pole). As can be seen from comparing $\sigma_L$ between Figs.~\ref{fig:iTL_binning}~($a$) and ($b$), the narrower peak, corresponding to a nearer pole, also challenges the reconstruction. This is consistent with the behavior observed in Refs.~\cite{Hansen:2017mnd, Bulava:2019kbi} in the context of obtaining amplitudes through solving inverse problems using Euclidean correlators.

\section{Conclusion \label{sec:conclusion}}

In this work, we have explored the prospects for extracting physical scattering amplitudes from Minkowski-signature correlation functions, calculated in a periodic one-dimensional spatial volume with extent $L$. This is relevant for future quantum simulations, in which real-time correlation functions can be accessed by preparing and time-evolving specific states. Defining finite-volume estimators for both hadronic and Compton amplitudes, we have shown how the formalism presented in Ref.~\cite{Briceno:2019opb} can be used to describe the finite-$L$ effects in terms of infinite-volume K matrices and related quantities. The relations hold for generic relativistic quantum theories to all orders in the interaction couplings and are exact up to terms scaling as $e^{-mL}$ where $m$ is the particle mass.

The relevant finite-volume expressions for both the Compton and hadronic estimators, denoted $\mathcal T_L$ and $\mathcal M_{L, [\mathcal J]}$ respectively, are summarized in Sec~\ref{sec:FV}. With an eye on the first quantum computations that will be performed, we restrict attention here to 1+1D systems and a single species of scalar particles. Following related work \cite{Hansen:2017mnd,Bulava:2019kbi}, we show that the physical amplitude can formally be extracted by introducing a complex energy, $E + i\epsilon$, and taking an ordered double-limit of ${L\to \infty}$ followed by ${\epsilon\to 0}$.

In Section~\ref{sec:numerical}, we investigate the required volumes to recover the infinite-volume Compton and hadronic amplitudes. We find that na\"ive analysis of a particular resonant system requires volumes of order $mL=\mathcal{O}(10^2)-\mathcal{O}(10^3)$, one to two orders of magnitude larger than volumes currently used in lattice calculations, to reduce finite-$L$ effects to the few-percent level. Here the situation may be worse in the lower-dimension theory as the density of $n$-particle states scales with $L^{d(n-1)}$, where $d$ is the number of spatial dimensions.

In order to overcome this issue, in Sec.~\ref{sec:numerical}(b) we use the fact that the infinite-volume Compton amplitude depends on fewer kinematic variables than its finite-volume analogue [see also Ref.~\cite{Hansen:2017mnd}]. 
Binning over redundant kinematic points, including multiple values of the total spatial momentum, we see that the resultant average converges faster to the desired observable. 
We stress that this is a very general approach, and that the formalism of Ref.~\cite{Briceno:2019opb} is only needed here to test the idea, but would not be required in a practical implementation. This means that the approach should naturally extend to multi-particle systems and other observables, for example the hadronic tensor. For a specific discussion of quantum computations of the latter see Ref.~\cite{Lamm:2019uyc}.

In future investigations it would be useful to better understand the optimal choice of $\epsilon(L)$, needed to extract the ordered double limit. This is a non-trivial issue as the best choice may depend on details of the theory, e.g.~resonance widths and the locations of multi-particle thresholds. One avenue that would be useful here is a better theoretical description of the $L$ and $\epsilon$ dependence, especially in the regime where many multi-particle channels are open. This could possibly be achieved, for example, by working at fixed order in $e^{- \epsilon L}$. [See also the discussion in Appendix \ref{app:averaging} as well as Refs.~\cite{Luscher:1985dn,Hansen:2020whp} for related ideas.] Finally, a more detailed understanding of the role of the spacetime dimension will be crucial as quantum computing develops beyond primitive 1+1D theories, towards 3+1D systems of direct phenomenological relevance.

\section{Acknowledgements}

RAB is supported in part by USDOE grant No. DE-AC05-06OR23177, 
under which Jefferson Science Associates, LLC, manages and operates Jefferson Lab.
RAB also acknowledges support from the USDOE Early Career award, contract de-sc0019229.
The authors would like to thank M.~Bruno, Z.~Davoudi, J.~Dudek, R.~Edwards, A.~Francis, D.~Grabowska, J.~Green, L.~Leskovec, and K.~Orginos for useful conservations. 

\appendix

\section{Finite-volume function and phase space in 1+1D \label{app:ps}}

In this appendix, we demonstrate the equivalence of Eqs.~\eqref{eq:Fdef} and \eqref{eq:Fcot} up to terms scaling as $e^{- m L}$. The first step is to rewrite \eqref{eq:Fdef}, by substituting $(P - k)^2 - m^2 = (E^\star - \omega_k^\star)^2 - k^{\star 2} - m^2 = E^\star(E^\star - 2 \omega_k^\star)$ and applying the Poisson summation formula
\begin{align}
F(E,\boldsymbol P, L)
&=
\frac{1}{2E^\star}\sum_{n\neq 0}\int\frac{dk }{2\pi} 
\frac{1}{2\omega_k } \frac{e^{inL k }}{E^\star-2\omega_k^\star } \,,
\label{eq:FA1}
\end{align}
where the sum runs over all nonzero integers. Here we have not explicitly displayed the $i \epsilon$ as this enters in a more complicated way. In particular the prescription uses $E + i \epsilon$ for the energy, and this is then passed into $\beta$, $\gamma$ as well as all quantities carrying a $\star$ superscript. Combining Eq.~\eqref{eq:FA1} with the Lorentz invariance of $dk/\omega_k$ and also substituting the boost relation, $k=\gamma (k^\star+\omega_k^\star \beta)$, then gives
\begin{align}
F(E,\boldsymbol P, L)
&=
\frac{1}{2E^\star}\sum_{n\neq 0}\int\frac{dk^\star}{2\pi} 
\frac{1}{2\omega_k^\star} \frac{e^{inL \gamma (k^\star+\omega_k^\star \beta) }}{E^\star-2\omega_k^\star } \,.
\end{align}
Next we multiply by 1 in the form $[E^\star+2\omega_k^\star]/[E^\star+2\omega_k^\star]$ and rearrange to reach
\begin{align}
F(E,\boldsymbol P, L)
&=
\frac{1}{2E^\star}\sum_{n\neq 0}\int\frac{dk^\star}{2\pi} 
e^{inL \gamma (k^\star+\omega_k^\star \beta) }
\frac{1}{2\omega_k^\star} \frac{4\omega_k^\star + (E^\star - 2\omega_k^\star )}{E^{\star2} - 4 \omega_k^{\star 2} } \,, \label{eq:Fmultone}\\
&=
\frac{1}{ 4 E^\star}\sum_{n\neq 0}\int\frac{dk^\star}{2\pi} 
\frac{e^{inL \gamma (k^\star+\omega_k^\star \beta) } }{q^{\star2} - k^{\star 2} } + \mathcal O(e^{- m L}) \label{eq:Fexpsup}\,, 
\end{align}
where in the second step we have used the fact that the $(E^\star - 2\omega_k^\star )$-term in the numerator of Eq.~\eqref{eq:Fmultone} cancels the pole and thus leads to exponentially suppressed $L$ dependence as shown in Eq.~\eqref{eq:Fexpsup}. Specifically, the $e^{- m L}$ scaling arises from the branch cuts within, $\omega_{k}^\star$ running from $k^\star = \pm i m$ to $\pm i \infty$.

At this stage, in order to evaluate the integral we need to examine the implicit factors of $\epsilon$ in $q^\star$ and $k^\star$. First note that the $\epsilon$-dependence of $k^\star$ arises from the relation 
\begin{align}
k^\star = \gamma (k - \omega_k \beta)= \frac{E + i \epsilon}{\sqrt{(E+i \epsilon)^2 - \boldsymbol P^2} } \bigg (k - \omega_k \frac{\boldsymbol P}{E + i \epsilon} \bigg) \,,
\end{align}
together with the fact that the original integral runs over real $k$. In other words, the integral over $k$ on the real axis is equivalent to integrating $k^\star$ along the contour defined by this expression. But it is straightforward to show that this can be deformed to the real line in $k^\star$ without changing the value of the integral. The next step is to make the $\epsilon$ dependence within $q^{\star 2}$ explicit. One finds $q^{\star 2}_{\epsilon} = q_0^{\star 2} + i \epsilon E_0/2 - \epsilon^2/4 $, i.e.~the poles are off the real line, exactly as for infinitesimal choices of $\epsilon$.

We are now ready to evaluate the integral, by closing the $k^\star$ contour in the upper or lower half of the complex plane. In doing so we encircle an isolated pole and a branch cut, but only the former contributes to the power-like $L$ dependence we are after. We find
\begin{align}
F(E,\boldsymbol P, L)&= - i
\frac{1 }{8 q^\star E^\star }
\sum_{n \neq 0}
\ \exp \! \Big [ {i L \gamma \big ( |n| q^\star + n \, \omega_q^\star \beta \big ) } \Big ]
+ \mathcal O(e^{- m L}) \,.
\label{eq:Fseries}
\end{align}
Summing the geometric series, with convergence guaranteed by the $i \epsilon$ prescription, we conclude 
\begin{align}
F(E,\boldsymbol P, L)&=
-i \rho(E^\star)
\left\{ - 2 +
\frac{1}{1- e^{i L \gamma (q^\star+\omega_q^\star \beta) }}
+
\frac{1}{1-e^{ i L \gamma (q^\star-\omega_q^\star \beta) }}
\right\} + \mathcal O(e^{- m L}),
\end{align}
which is equivalent to Eq.~\eqref{eq:Fcot}. Note that (for real $E^\star$) ${\rm Im}\, F(E,\boldsymbol P, L) = \rho(E^\star)$.

\section{Boost averaging \label{app:averaging}}

Starting with Eq.~\eqref{eq:Fseries}, $F(E,\boldsymbol P, L)$ can be conveniently written as 
\begin{align}
F(E_\epsilon,\boldsymbol P, L) = -2i\rho(E^\star_\epsilon) \sum_{n > 0} 
e^{i n L \gamma_\epsilon q^{\star}_\epsilon } \cos \big ( n L \gamma_\epsilon \omega_{q,\epsilon}^\star \beta_\epsilon \big ) \,,
\end{align}
where we have combined the $n>0$ and $n<0$ pairs into the cosines. Here we have also dropped the $\mathcal O(e^{- mL})$ and will neglect this term throughout this section, as we have also done in the main text. In addition, the $\epsilon$ has been made explicit in all quantities that depend on this parameter. This expression can be simplified by substituting $L \gamma_{\epsilon} \omega_{q,\epsilon}^\star \beta_{\epsilon} = L \gamma E_{\epsilon}^\star (\boldsymbol P/E_{\epsilon}) /2 = L \boldsymbol P/2 = \pi \boldsymbol d$ with $\boldsymbol P = 2 \pi \boldsymbol d/L$, implying
\begin{align}
F(E_{\epsilon},\boldsymbol P, L) = -2i\rho(E^\star_{\epsilon}) \sum_{n > 0} (-1)^{n \boldsymbol d} \,
\left(e^{ i L \gamma_{\epsilon} q_{\epsilon}^\star }\right)^n \, .
\label{eq:FseriesII}
\end{align}
Thus, the sum exhibits a very simple phase oscillation that holds exactly, even at non-infinitesimal values of $\epsilon$.

To reduce further, we use the fact that $\text{Im}[\gamma_\epsilon q^\star_\epsilon] > 0$ implying $\big (e^{- \epsilon L \text{Im}[\gamma_\epsilon q^\star_\epsilon]} \big)^n$ is suppressed for larger values of $n$. In addition, although we are working with non-infinitesimal values of $\epsilon/m, 1/(mL)$, we do take the parameters significantly smaller than one, in order to achieve an approach towards the infinite-volume amplitudes. This motivates an expansion of $F$ based in the power-counting scheme $\{\epsilon/m,\ 1/(mL), \ e^{- L \text{Im}[\gamma_\epsilon q^\star_\epsilon]} \}= \mathcal O(\delta)$. The leading-order expression is given by
\begin{align}
F (E_{\epsilon}^\star, \boldsymbol P, L ) & =
-2i\rho(E^\star_0) \left(e^{- L \epsilon} \right)^{\alpha_0} e^{i L \gamma_0 q_0^{\star} } (-1)^{\boldsymbol d} + \mathcal O(\delta^2) \,,
\label{eq:Fseriesapprox}
\end{align}
where $\alpha_0$ is given by
\begin{align}
\alpha_0 &= \frac{\partial (\gamma_{\epsilon} q^\star_{\epsilon})}{\partial (i \epsilon)} \bigg \vert_{\epsilon=0} = \frac{E_0^{\star4}+4 m^2 \boldsymbol P^2 }{4 q_0^{\star}  E_0^{\star 3} }  \,.
\end{align}

We now turn to the boost averaged $F$, denoted by $\widetilde{F}$. Substituting our approximate form gives
\begin{align}
\widetilde F(E_\epsilon, L) &\equiv \frac{1}{N_d} \sum_{\boldsymbol d=0}^{N_d-1} F(E_{\epsilon},\boldsymbol P, L)\label{eq:Ftilde}
= -2i\rho(E^\star_0) \frac{1}{N_d} \sum_{\boldsymbol d=0}^{N_d-1} 
\left(e^{- L \epsilon} \right)^{\alpha_0} 
e^{i L q_0^{\star} \gamma_0} (-1)^{\boldsymbol d} + \mathcal O(\delta^2) \,,
\end{align}
where $N_d$ is the number of boosts. Observe that the sum appearing in $\widetilde F(E, L)$ does not scale as $N_d$, due to the alternating sign of $(-1)^{\boldsymbol d}$. As a result the boost averaged value of $F$, and thus also the $L$ dependence of the finite-volume amplitudes discussed in Secs.~\ref{sec:FV} and \ref{sec:numerical}, is suppressed by $1/N_{d}$.

\begin{figure}[t!]
\begin{center}
\includegraphics[width=1\textwidth]{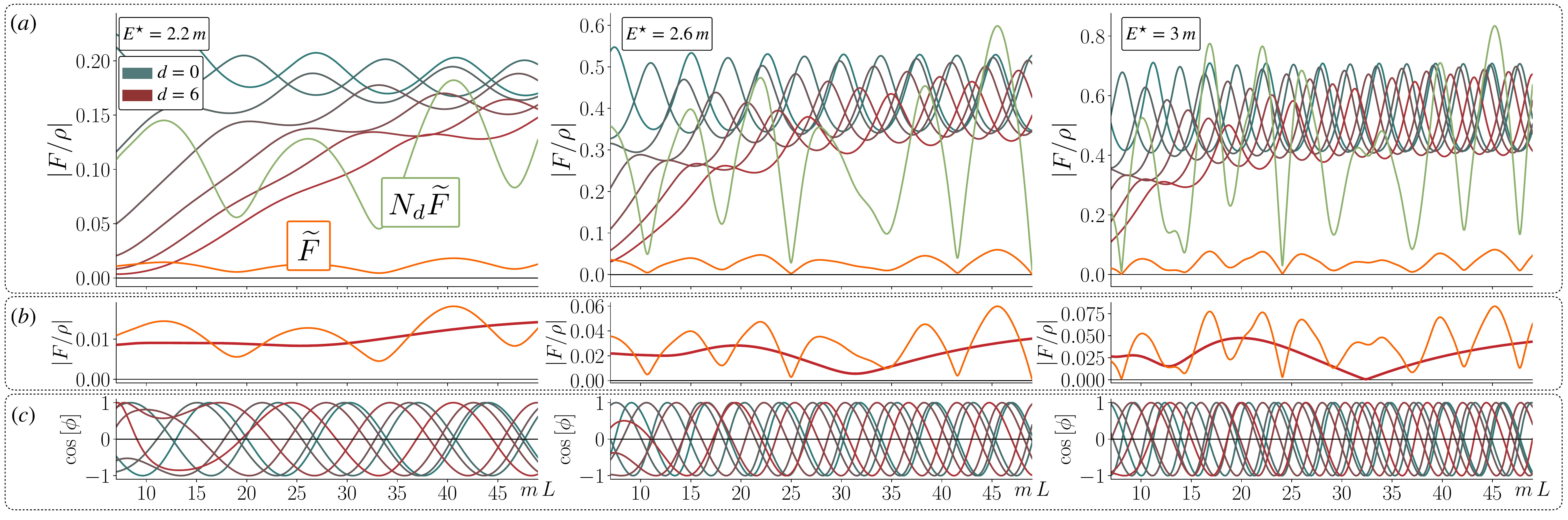}
\caption{$(a)$ Finite-volume function $F(E + i \epsilon, \boldsymbol P, L)$, for seven different momenta ($\boldsymbol d = 0,1,\dots,6$, with $\boldsymbol P = 2 \pi \boldsymbol d/L$; colored red to teal). We additionally plot the boost-averaged function $\widetilde{F}$ (orange line) and the product $N_d \widetilde{F}$ (light-green). $(b)$ Comparison of the average function, $\widetilde{F}$, with the approximation given by Eq.~\eqref{eq:Ftilde}. $(c)$ Complex phase for each individual $F$ [$\phi = \arctan(\text{Im}F/\text{Re}F)$], for the seven momenta plotted in $(a)$. We set $\epsilon L = 1$ in all plots.\label{fig:F_ave} 
}
\end{center}
\end{figure}

In Fig.~\ref{fig:F_ave}(a) we plot the function $F(E + i \epsilon, \boldsymbol P, L)$ for the first seven values of total momenta, $\boldsymbol d = 0$ to $6$, together with the sum and the average over these. Note that the characteristic magnitude of $F$ is largely independent of $\boldsymbol d$, as predicted by Eq.~\eqref{eq:Fseriesapprox}. Moreover, the magnitude of the sum over the momentum set, given by $N_d |\widetilde{F}(E^\star,L)|$, is also of the order of $|F|$, due to the destructive interference of the phases. 
This, in turn, implies that the boost-averaged  quantity, $\widetilde{F}$, is suppressed by the factor $1/N_d$ in Eq.~\eqref{eq:Ftilde}. Figure~\ref{fig:F_ave}(b) shows the asymptotic approximation of $\widetilde{F}$ given by Eq.~\eqref{eq:Ftilde} (red), compared to the exact average (orange). The orange curve oscillates around its asymptotic approximation, indicating that the power-counting scheme detailed above gives a reasonable description for the values of $\epsilon$, $L$ and $E$ considered here. Finally, in Fig.~\ref{fig:F_ave}(c) we plot the complex phase of the function $F(E + i \epsilon, \boldsymbol P, L)$ for each value of $\boldsymbol d$, again to give a sense of the destructive interference between different values of total spatial momenta.

\bibliography{bibi}
\end{document}